\documentclass[10pt,sigconf]{acmart}
\usepackage{balance}
\usepackage{xspace}
\usepackage{graphicx}
\usepackage{amsmath}
\usepackage{textcomp}
\usepackage{hyphenat}
\usepackage{float}
\usepackage[T1]{fontenc}
\usepackage[utf8]{inputenc}
\usepackage[labelformat=simple]{subcaption}
\usepackage{array}
\usepackage{tabularx}
\usepackage{booktabs}

\newcolumntype{P}[1]{>{\centering\arraybackslash}p{#1}}
\newcolumntype{M}[1]{>{\centering\arraybackslash}m{#1}}

\DeclareMathOperator*{\argmax}{arg\,max}

\newcommand{\ie}{{\it i.e.}\xspace}
\newcommand{\eg}{{\it e.g.}\xspace}

\newcommand{\parahead}[1]{\vspace*{1ex plus 0.25ex minus 0.25ex}\noindent %
  {\bfseries #1.}}
\newcommand{\parabreak}{\vspace*{1.5ex}\noindent}
 
\graphicspath{{../}{figures/}}

\newcommand{\systemname}{mD\hyp{}Track\xspace}

\begin{document}
\begin{titlepage}
\center
\newcommand{\HRule}{\rule{\linewidth}{0.5mm}} % Defines a new command for the horizontal lines, change thickness here
\textsc{\LARGE  }\\[4.5cm]

\HRule \\[0.4cm]
{ \huge \bfseries \systemname: Leveraging Multi-Dimensionality in Passive Indoor Wi-Fi Tracking}\\[0.4cm] % Title of your document
\HRule \\[0.5cm]
\huge Yaxiong \textsc{Xie}, Jie \textsc{Xiong}, Mo \textsc{Li}, Kyle \textsc{Jamieson}
\textsc{\LARGE  }\\[1.5cm]
\textit{\LARGE This work is a pre-print version to appear at MobiCom 2019.}\\[0.5cm] % Minor heading such as course title

\end{titlepage}
%\newcommand{\TITLE}{Breaking the resolution limit of localization with\\ multi-dimensional information}
%  \copyrightyear{2018} 
%  \acmYear{2018} 
%  \setcopyright{acmcopyright}
%  \acmConference[MobiCom '18]{The 24th Annual International Conference on Mobile Computing and Networking}{October 29-November 2, 2018}{New Delhi, India}
%  \acmBooktitle{The 24th Annual International Conference on Mobile Computing and Networking (MobiCom '18), October 29-November 2, 2018, New Delhi, India}
%  \acmPrice{15.00}
%  \acmDOI{10.1145/3241539.3241572}
%  \acmISBN{978-1-4503-5903-0/18/10}
\setcopyright{none}
\title{\systemname: Leveraging Multi-Dimensionality in Passive Indoor Wi-Fi Tracking}
 
\author{Yaxiong Xie}
\affiliation{%
  \institution{Nanyang Technological University}
}

\author{Jie Xiong}
\affiliation{%
  \institution{UMass Amherst}
}

\author{Mo Li}
\affiliation{%
  \institution{Nanyang Technological University}
}

\author{Kyle Jamieson}
\affiliation{%
  \institution{Princeton University}
}

\begin{abstract}
Wi-Fi localization and tracking face accuracy limitations dictated by
antenna count (for angle-of-arrival methods) and frequency bandwidth
(for time-of-arrival methods).  This paper presents \systemname, a
device-free Wi-Fi tracking system capable of jointly fusing
information from as many dimensions as possible to overcome the
resolution limit of each individual dimension.  Through a novel path
separation algorithm, \systemname can resolve multipath at a much
finer-grained resolution, isolating signals reflected off targets of
interest.  \systemname can localize human
passively at a high accuracy with just a single Wi-Fi transceiver
pair. \systemname also introduces novel methods to greatly streamline its estimation algorithms, achieving real-time operation.  We implement \systemname on both WARP and cheap
off-the-shelf commodity Wi-Fi hardware, and evaluate its performance
in different indoor environments.  

\end{abstract}
\begin{CCSXML}
<ccs2012>
	<concept>
	<concept_id>10003033.10003058.10003065</concept_id>
	<concept_desc>Networks~Wireless access points, base stations and infrastructure</concept_desc>
	<concept_significance>500</concept_significance>
	</concept>
</ccs2012>
\end{CCSXML}
\ccsdesc[500]{Networks~Wireless access points, base stations and infrastructure}

\keywords{Localization, AoA, ToF, Doppler, Multi-dimensional Estimation, Super-resolution Algorithm, Diversity}

\maketitle

\section{Introduction}
\label{s:intro}

Passive motion tracking without any device carried by or attached to a
person has been an exciting area of recent interest, with important
applications including security surveillance~\cite{PhyCloak}, elderly
care~\cite{WiTrack1}, and retail business~\cite{ShopMiner}.  Diverse
technologies have been proposed for localization and tracking
including ultrasound~\cite{cricket-mobicom00}, infrared~\cite{bat,
active-badge}, cameras~\cite{hile-ieeegraphics08}, and LED visible
light~\cite{epsilon-nsdi14, pharos-hotnets13,VLCsense}. Among these
technologies, Wi-Fi based systems stand out as particularly promising
due to the pervasive availability of Wi-Fi access points~(APs).  For
device-free passive tracking, Wi-Fi based systems rely on signal
reflections off targets to extract essential motion and location
information. By its nature, passive tracking is more challenging than
localization of active wireless transmitters, because reflected
signals of interest are typically orders of magnitude weaker than the
direct\hyp{}path signal, and are typically superimposed with the
strong direct\hyp{}path signal as well as signals reflected from
walls, furniture, and other nearby clutter.  How to accurately resolve
and identify the weak signal reflected off a target of interest
becomes a major challenge for these systems.

\begin{figure}
	\centering
	\includegraphics[width=0.8\linewidth]{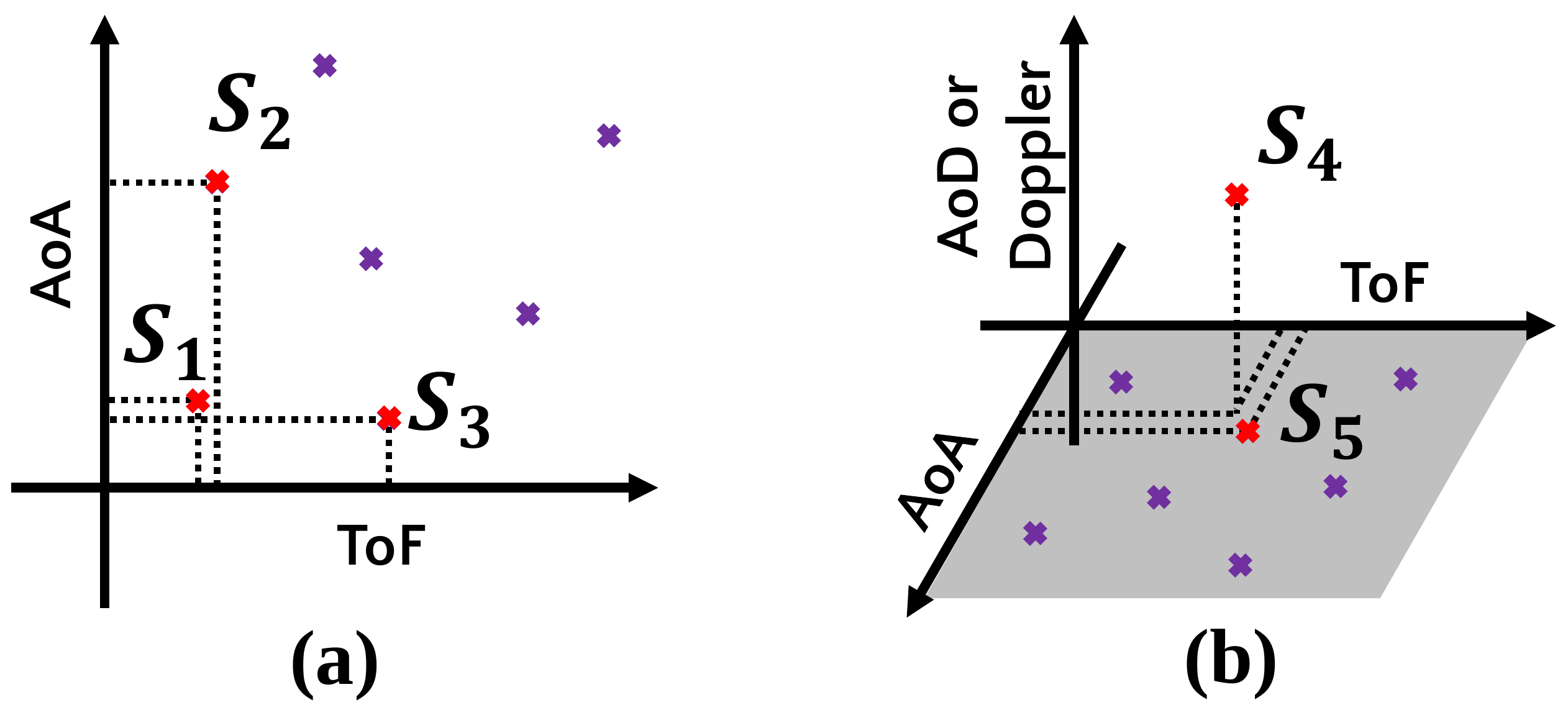}
  \caption{Joint estimation in the {\bfseries (a)} Time-of-Flight (ToF)
  and Angle-of-Arrival (AoA) dimensions, {\bfseries (b)} ToF,
  AoA, and Doppler shift or Angle-of-Departure (AoD) dimensions, which
  can separate incoming signals more effectively.}
	\label{f:jointEstimation}
\end{figure}

Recent progress in this area has explored many ways of extracting
different \emph{parameters} of the wireless signal, such as
angle\hyp{}of\hyp{}arrival (AoA) and time\hyp{}of\hyp{}flight (ToF)
\cite{ArrayTrack, Phaser, CUPID} for localization and tracking. These
approaches rely on accurately estimating the AoA or ToF of each signal
path, and so when multiple paths have similar AoAs or ToFs, these
systems face fundamental difficulties resolving the paths to obtain
accurate parameter estimates.  

Resolvability is determined by the
number of antennas (for AoA) and the transmission frequency bandwidth
(for ToF), respectively. So a straightforward way to improve AoA
resolution is to increase the number of antennas and radios, which
results in higher hardware cost.  Improving ToF resolution is harder,
as Wi\hyp{}Fi standards fix channel bandwidth. Recent attempts to
overcome these limitations include creating virtual antenna arrays by
physically moving the antenna and combining adjacent channels to form
a larger bandwidth with channel hopping~\cite{Yaxiong-mobicom15,
tonetrack-mobicom15, Ubicarse, SWAN}. However, these methods may impose
constraints on ongoing data communication that we seek to avoid here,
such as the use of a wider bandwidth channel in a situation where use
of a narrower bandwidth channel would be more favorable from a
communications standpoint.  While other recent systems~\cite{WiDeo,
SpotFi} jointly estimate AoA and ToF, they are inherently limited, by
design, to the two signal dimensions they can explore, and have high
computational complexity, so are not easily scalable to higher
dimensions. 

In this work, rather than adding more antennas or increasing channel
bandwidth, we explore more dimensions of the wireless
signal itself.  We illustrate the intuition behind this idea in
Fig.~\ref{f:jointEstimation}.  Three signals arrive at the receiver
simultaneously. Signals $S_1$ and $S_2$ are close in time and
therefore cannot be resolved by employing ToF. However, $S_1$ and
$S_2$ can be easily resolved with AoA, since their AoA difference is
large. Similarly, signals $S_1$ and $S_3$ cannot be separated with
AoA, but are resolvable with ToF. This concept extends to higher
dimensions as shown in Fig.~\ref{f:jointEstimation}(b), where the
receiver jointly estimates ToF, AoA, and a third signal parameter 
(Doppler shift or Angle-of-Departure--AoD). Here $S_4$ and
$S_5$ are close to each other in both AoA and ToF, but since $S_4$ is
a signal from a moving source, it exhibits a non-zero Doppler shift
that separates it from $S_5$, a reflected signal from a static object
with zero Doppler shift.  There is thus an opportunity
to improve signal resolvability by jointly exploiting information from
more signal dimensions, without changing the resolution limit of any
individual dimension.

\vspace*{0.5ex plus 0.25ex minus 0.25ex}\noindent This paper leverages
the foregoing opportunity, describing \emph{multi\hyp{}Dimensional
Track} (\emph{\systemname}), a passive Wi-Fi tracking system that
fuses information from multiple signal dimensions, significantly
improving resolvability without requiring a wider frequency bandwidth
or a larger number of antennas.  \systemname jointly estimates
multi-dimensional parameters simultaneously so that the respective
parameters corresponding to a single path can be easily associated
with that path, improving passive localization and tracking of the
motions of multiple targets simultaneously with only a single
transmitter\hyp{}receiver pair.
\systemname makes the following contributions:

\noindent{\bfseries 1.\enspace{}Multi-dimensional signal estimator.}
\systemname{} introduces a signal processing structure (the
\emph{multi\hyp{}dimensional estimator}, shown in
Fig.~\ref{f:MF_design}), that combines sources of information from all
available signal parameters into a single metric.  This structure
serves as the building block in our next algorithmic contribution.

\noindent{\bfseries 2.\enspace{}Iterative path parameter refinement.}
Separating signals of close\hyp{}by paths is challenging:\ reflected
signals are much weaker compared to the direct path signal, so it is
difficult to accurately estimate their parameters in the presence of
interference from a strong direct path.  To separate incoming
signals, \systemname employs an iterative path parameter refinement
method during which the signals are iteratively re\hyp{}estimated,
more accurately reconstructed, and then subtracted from the received
signal with refined parameters in multiple rounds of estimation.  This
design recalls the structure of the Turbo Decoder \cite{turbo} and
also can be shown to be an expectation maximization estimation
algorithm.

\noindent{\bfseries 3.\enspace{}Bounding computation for real time
operation.} While multi\hyp{}dimensional joint estimation helps to
improve signal resolvability and parameter estimation accuracy, the
computation required by the joint estimator increases exponentially
with the number of signal dimensions, making the required amount of
computation intractable for a
real\hyp{}time system design.  To address this challenge, we design a
linear\hyp{}time estimator by exploiting a coordinate descent method
together with the expectation maximization algorithm to reduce
computational complexity significantly, making the design practical
for real\hyp{}time operation~(\S\ref{s:computationOverhead}).

We implement \systemname on both WARP and commercial
off\hyp{}the\hyp{}shelf (COTS) Wi-Fi APs. Our experimental evaluation
begins with a sensitivity analysis that analyses each signal
parameter, measuring the relative ability of different parameters to
resolve signals.  Head\hyp{}to\hyp{}head comparisons in parameter
estimation and passive localization demonstrate $3.5\times$ accuracy
improvements over the state\hyp{}of\hyp{}the\hyp{}art SpotFi
\cite{SpotFi} system.  Further experiments measure the effect of
adding another signal dimension (Doppler shift) to \systemname,
showing approximately a $3\times$ accuracy improvement, in contrast
with a marginal 20\% improvement from doubling the frequency
bandwidth.  Experiments show that with the
three antennas available on the
COTS Wi-Fi card, \systemname can resolve more than 10 signals
and estimate the parameters of each signal path accurately.

% \paragraph{Roadmap.}  The rest of this paper is structured as follows. Section~\S\ref{s:w_channel} gives the wireless channel model. Section~\S\ref{s:para} presents our design of parameter estimation algorithm followed by the channel measurement and error handling in \S\ref{s:chMea}. Our implementation is presented in \S\ref{s:impl}. Our experimental evaluation is presented in \S\ref{s:eval}, related work in \S\ref{s:related}, followed by a conclusion in \S\ref{s:concl}.

\section{The Wireless Channel}
\label{s:w_channel}
 
\begin{figure}[htb]
\centering
\includegraphics[width=0.8\linewidth]{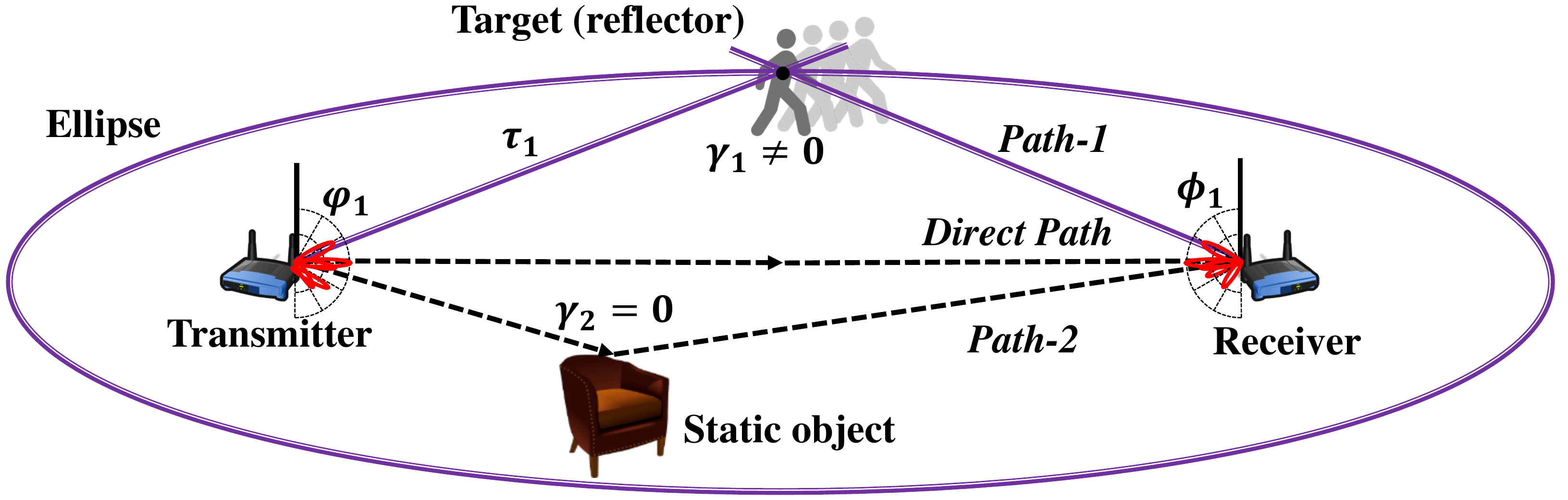}
\caption{The multi-dimensional parameters of signal paths related to
location and motion tracking.}
\label{f:pathParameter}
\vspace{-1mm}
\end{figure}

Localization or motion tracking of a target relies on separating superimposed signals and accurately estimating each signal’s parameters. A wireless signal is characterized by
multi-dimensional parameters, each parameter providing us a piece of location or motion information about the target.
Fig.~\ref{f:pathParameter} summarizes the parameters that can be retrieved from a wireless signal reflected from a human target. We only consider the path with one or less reflections, since the signal experiencing two or more reflections during the transmission, has extremely low signal strength and can hardly be captured by wireless receiver.

% A detailed description of each parameter is as follows.}

\subsection{Parameters of a signal path}

\parahead{Time of flight ($\boldsymbol\tau$)} The propagation time the signal takes to travel along a particular path from the transmitter to the receiver is referred as the {\itshape time of flight}~(ToF) $\tau$. 
As Fig.~\ref{f:pathParameter} depicts, ToF estimation of a reflected signal defines an ellipse (with the transmitter and receiver as the two focal points) where the reflector is located.  The resolution of ToF estimation is inversely proportional to the channel bandwidth~\cite{Yaxiong-mobicom15}.

\parahead{Angle of arrival ($\boldsymbol\phi$) and angle of departure ($\varphi$)} The {\itshape angle of arrival} (AoA) $\phi$ indicates the direction of the signal arriving at the receiver, and the \emph{angle of departure} (AoD) $\varphi$ indicates the direction of the signal departure from the transmitter, as shown in Fig.~\ref{f:pathParameter}.  The number of antennas at sender/receiver determines the resolution of the AoA/AoD estimates, respectively. 

\parahead{Doppler shift ($\boldsymbol\gamma$)} Movement of the transmitter, receiver, or reflectors all introduce frequency shifts to the carrier frequency of the signal which is referred as {\itshape Doppler shift} $\gamma$. 
Doppler resolution is related to the observation interval: the longer the interval, the finer the resolution.

\parahead{Complex attenuation ($\boldsymbol\alpha$)} The signal is attenuated by $\alpha$ when propagating from the sender towards the receiver.

\subsection{Wireless signal model} 
\label{s:sig_model}
\begin{figure*}[t]
	\centering
	\includegraphics[width=0.94\linewidth]{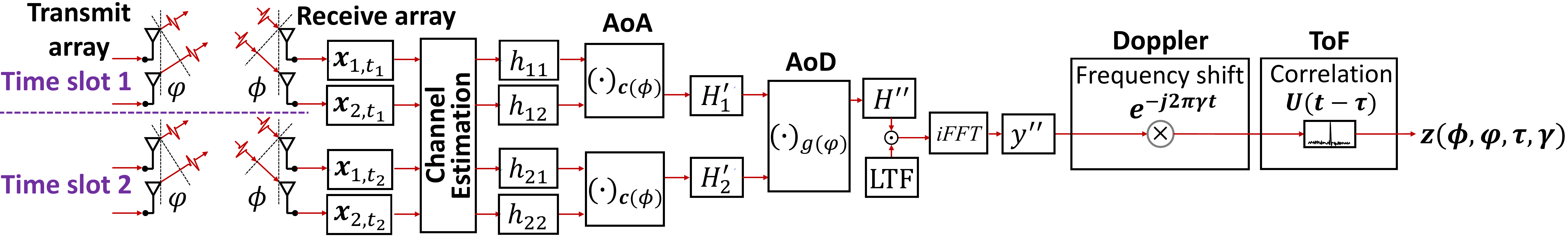}
	\caption{\systemname's four\hyp{}dimensional estimator
		that estimates parameters AoA, AoD, Doppler, and ToF of a wireless
		signal as it propagates along a single path. Receive and transmit antenna steering vector $\mathbf{c}(\phi)$ and $\mathbf{g}(\varphi)$ are defined in Section~\ref{s:sig_model}. }
	\label{f:MF_design}
	\vspace{-3mm}
\end{figure*}

The \systemname transmitter has an array of $N$ antennas, and 
receiver has an array of $M$ antennas. They are linear arrays with a
uniform spacing $d$ between adjacent antennas, as shown in
Fig.~\ref{f:MF_design}. If we denote the transmitted signal as
$U(t)=[u_1(t),u_2(t),\dots,u_N(t)]$, then we can use the above signal parameters to express the signal reaches the receiver through a single path as follows: 
\begin{eqnarray}
\label{eqn:s_signal}
\mathbf{s}(t;\upsilon) =  \alpha 
  e^{j2\pi \gamma t} 
  \mathbf{c}(\phi)\mathbf{g}(\varphi)^T \mathbf{U}(t-\tau),
\end{eqnarray}
where $\mathbf{\upsilon}=[\phi,\varphi,\tau,\gamma,\alpha]^T$ is the
\textit{parameter vector} containing the parameters that
characterize the signal.  The $N$\hyp{}element {\itshape transmit
array steering vector} $\mathbf{g}(\varphi)$ characterizes the phase
relationship of the signal coming out of the $N$ transmitting antennas
while the $M$-element \emph{receive array steering vector}
$\mathbf{c}(\phi)$ characterizes the phase relationship of the signal
arriving at the $M$ receiving antennas~\cite{ArrayTrack}. The signal received at the
receive antenna due to this single path is then: 
\begin{eqnarray}
\label{eqn:noisySignal}
\mathbf{y}(t) = \mathbf{s}(t) + \mathbf{W}(t),
\end{eqnarray}
where $\mathbf{W}(t)=[w_1(t),w_2(t),\dots,w_M(t)]^T$ is $M$\hyp{}dimensional
complex Gaussian noise capturing the background noise.\footnote{We use $s(t)$ and $s(t;\upsilon)$ interchangeably throughout the paper.}
\section{Parameter Estimation}
\label{s:para}

We first describe the design of \systemname's parameter estimation
algorithm in Section~\ref{s:multiEst} in a simplified scenario where
there is only one path in the environment, and describe in detail how
different signal parameters (ToF, AoA, AoD, Doppler shift and
attenuation) of that single path can be estimated. Then in
Section~\ref{s:resovMultipath} we introduce our joint estimation
method to handle multiple signals arriving through different
propagation paths.  Section~\ref{s:computationOverhead} discusses our
approach to make computational complexity tractable.
\subsection{Multi-dimensional estimator}
\label{s:multiEst}\label{s:design:mde}

This section first considers an environment where there is only one path from sender to receiver. We assume perfect transceivers without phase offsets across radio chains. Such an assumption is justified in Section~\ref{s:ChanM_error} with phase calibration. Our design factors into different modules, each corresponding to one of the above parameters. Multiple modules are employed together to jointly estimate all the parameters.

\begin{figure}[h]
	\centering
	\includegraphics[width=0.92\linewidth]{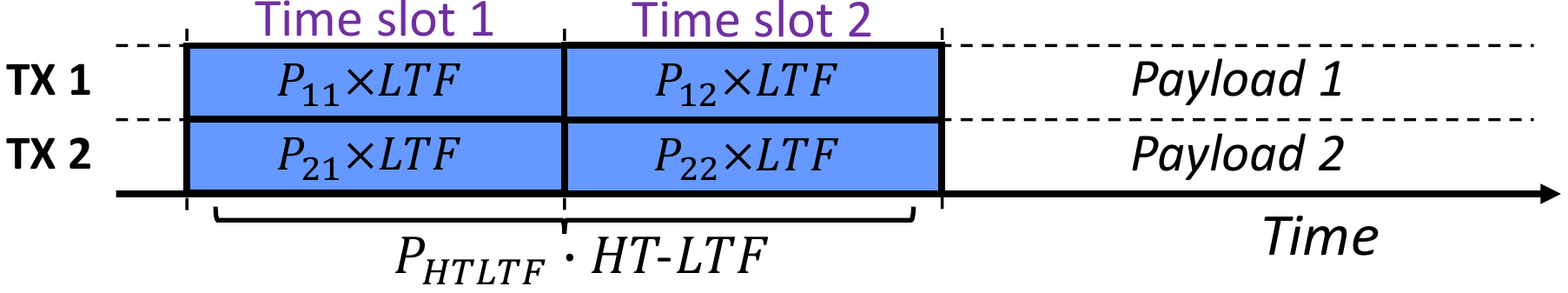}
	\caption{802.11n (High Throughput) preamble.}
	\label{f:AoD_preamble}
	\vspace{-3mm}
\end{figure}

\parahead{Channel estimation} Before estimating the parameters, we need to estimate the wireless channels $h_{i,j,k}$ of subcarrier $k$ between each transmit antenna $i$ and receive antenna $j$. For explanatory purposes, we will consider a running example with $N$ subcarriers and two transmit and two receive antennas as shown in Fig.~\ref{f:MF_design}. The task is to estimate:
% we need to acquire signal transmitted by two antennas in the transmitting array in two consequective time slots, or equivalently the channels 
\begin{eqnarray}
\label{eqn:H_mat}
\mathbf{H}_k =  \begin{bmatrix}h_{11,k} & h_{12,k}\\
                        h_{21,k} & h_{22,k}\end{bmatrix}.
\end{eqnarray}
for all subcarriers $k=[1,2,\dots,N]$.
% between two transmitting and two receiving antennas, just as shown in Figure~\ref{f:MF_design}.
To avoid sudden signal level changes,
% The silent antenna, however, introduces a sudden signal decrease and causes problem to the AGC control. Therefore, 
the 802.11n standard~\cite{80211n} multiplies the preamble or \textit{High Throughput Long Training Field}~(HT-LTF\footnote{We use HT-LTF and LTF interchangeably throughout the paper.}) with the \textit{HT-LTF mapping matrix}:
\begin{eqnarray}
\label{eqn:P_HTLTF}
\mathbf{P}_{HTLTF} = \begin{bmatrix}1 & -1\\
                        1 & 1\end{bmatrix} ,
\end{eqnarray}
whose columns correspond to each of the two time slots and whose rows correspond to each of the two transmit antennas. This results in the frequency domain transmitted signal $\mathbf{P_{HTLTF}}\times\mathbf{LTF}(k)$, as shown in Fig.~\ref{f:AoD_preamble}.
When both antennas transmit simultaneously, the received frequency domain signal from subcarrier $k$ at the two receive antennas during the two time slots $t_1$ and $t_2$ is given by:
\begin{eqnarray}
\label{eqn:RX_signal}
\begin{bmatrix}
\mathbf{x}_{1,k,t_1} & \mathbf{x}_{1,k,t_2}\\
\mathbf{x}_{2,k,t_1} & \mathbf{x}_{2,k,t_2}
\end{bmatrix}= \begin{bmatrix}h_{11,k}+ h_{21,k} & -h_{11,k}+h_{21,k}\\
                        h_{12,k}+h_{22,k} & -h_{12,k}+h_{22,k}\end{bmatrix} \times \mathbf{LTF}(k) \nonumber
\end{eqnarray}
Since $\mathbf{P}_{HTLTF}$ is known and constant, the receiver estimates the wireless channel by multiplying the received signal with $\mathbf{P}_{HTLTF}^*$, and obtains:
\begin{eqnarray}
\label{eqn:chan_est_mat}
\begin{bmatrix}
\mathbf{x}_{1,k,t_1} & \mathbf{x}_{1,k,t_2}\\
\mathbf{x}_{2,k,t_1} & \mathbf{x}_{2,k,t_2}
\end{bmatrix}\times \mathbf{P}_{HTLTF}^* 
= 2\times\begin{bmatrix}h_{11,k} & h_{21,k}\\
        h_{12,k} & h_{22,k}  
        \end{bmatrix} \times \mathbf{LTF}(k). \nonumber
\end{eqnarray}
Via the above process, the receiver decouples the two simultaneously transmitted preambles to estimate the wireless channel. We note that 802.11's cyclic time delay across different antennas, introduces known linear phase shifts across subcarriers on $h_{21,k}$ and $h_{22,k}$, which are removed by \systemname after extracting the channel estimates.

\parahead{AoA estimator} The AoA estimator, \textit{i.e.,} the \textit{AoA} box in Fig.~\ref{f:MF_design}, is implemented by multiplying the estimated channel $\mathbf{H}$ with the receive antenna array steering vector $\mathbf{c}(\phi)$ defined in Section~\ref{s:sig_model}. Specifically, on subcarrier $k$, we obtain:
\begin{eqnarray}
\label{eqn:AoA_multiply}
\begin{bmatrix}
h'_{1,k}(\phi)\\
h'_{2,k}(\phi)
\end{bmatrix} =  \begin{bmatrix}
h_{11,k} & h_{21,k}\\
h_{21,k} & h_{22,k}
\end{bmatrix}\mathbf{c}^*(\phi).
\end{eqnarray}
For the wideband Wi-Fi channel with $N$ subcarriers, we apply Eq.~\ref{eqn:AoA_multiply} to the estimated channels of all subcarriers and obtain the combined channel:
\begin{eqnarray}
\label{eqn:WideChannel}
% \phi  = 
\mathbf{H}'(\phi) = \begin{bmatrix}
\mathbf{H}'_1(\phi)\\
\mathbf{H}'_2(\phi)
\end{bmatrix} = \begin{bmatrix}
h'_{1,1}(\phi) & \dots & h'_{1,k}(\phi) & \dots & h'_{1,N}(\phi) \\
h'_{2,1}(\phi) & \dots & h'_{2,k}(\phi) & \dots & h'_{2,N}(\phi) 
\end{bmatrix}
\end{eqnarray}
Differing AoAs $\phi$ yield differing steering vectors, so once 
the correct $\phi$ is applied to the channel $\mathbf{H}$, the channel between the transmit antenna and the two receive antennas align and add constructively, so the strength of the AoA estimator output is maximized.
Therefore, we estimate the AoA by searching for the $\phi^*$ that maximizes the sum power of the combined channel: 
\begin{eqnarray}
\label{eqn:AoA_est}
% \phi  = 
\phi^* = \argmax_{\phi} \sum_{i=1}^2\sum_{k=1}^N\left \|h'_{i,k}(\phi) \right \|^2.
%       &=& M\exp\left\{j2\pi \gamma\right\} U(t-\tau) + n
\end{eqnarray}
where $h'_{i,k}(\phi)$ is the combined channel of the $k^{th}$ subcarrier on the $i^{th}$ antenna.

\parahead{AoD Estimator} 
When received, the second transmit antenna's signal travels an additional distance to reach the receive array, and hence introduces an extra phase shift.
The \systemname AoD estimator takes $\mathbf{H}'(\phi^*)$ as input and correlates this matrix with the transmit antenna array steering vector $\mathbf{g}(\varphi)$ defined in Section~\ref{s:sig_model}:
\begin{eqnarray}
\label{eqn:AoD_multiply}
\mathbf{H}''(\varphi;\phi^*) = \mathbf{g}^H(\varphi) \mathbf{H}'(\phi^*),
\end{eqnarray}
where the combined channel $\mathbf{H}''(\varphi;\phi^*)$ is a $1\times N$ vector. We estimate the AoD by searching for the $\varphi^*$ that maximizes the sum power of the combined channel: 
\begin{eqnarray}
\label{eqn:bf_signal}
\varphi^* = \argmax_{\varphi} \sum_{k=1}^N \left \| h''_k(\varphi;\phi^*) \right\|^2,
\end{eqnarray}
where $h''_k(\varphi;\phi^*)$ is the $k$-th element of $\mathbf{H}''(\varphi;\phi^*)$, \textit{i.e.,} the channel of the $k$-th subcarrier.
As with the AoA estimator, once the AoD guess is correct, the phase differences due to AoD are removed. The channels thus constructively combine, maximizing the output magnitude.

\parahead{Doppler and ToF Estimator} We reconstruct the received time domain signal using the channel matrix $\mathbf{H}''(\varphi^*,\phi^*)$ and the known transmitted frequency domain $\mathbf{LTF}$ by \textit{iFFT}:
\begin{eqnarray}
\label{eqn:bf_signal}
\mathbf{y}''(t;\phi^*,\varphi^*) = \mathcal{F}^{-1}\{\mathbf{H}''(\varphi^*,\phi^*) \odot \mathbf{LTF}\},
\end{eqnarray}
where the operator $\odot$ is the Hadamard product, \textit{i.e.,} element wise multiplication of matrix.
The signal $\mathbf{y}''(t;\phi^*,\varphi^*)$ that arrives at the Doppler and ToF estimators is a frequency-shifted and delayed version of the transmitted signal $\mathbf{U}(t)$. 
To estimate that frequency shift and the delay, our approach is to cancel the frequency shift and reverse the delay of the signal and then correlate it with the transmitted signal $\mathbf{U}(t)$. If the correct frequency shift is canceled and correct delay is reversed, there will be a peak in the correlation result.

Specifically, the Doppler shift causes a $2\pi \gamma t$ phase shift to the received time domain signal according to Eq.~\ref{eqn:s_signal}. We therefore cancel such a frequency shift by multiplying the signal by $e^{-j2\pi \gamma t}$. To reverse the delay and obtain $\mathbf{y}''(t+\tau;\phi^*,\varphi^*)$ is practically difficult since $\mathbf{y}''(t;\phi^*,\varphi^*)$ is the received signal. Instead of reversing the delay, we correlate the received signal with the delayed version of transmitted signal $\mathbf{U}(t-\tau)$. For a given $\phi^*$ and $\varphi^*$ the correlation is computed as:
\begin{eqnarray}
\label{eqn:z_tau_gamma}
z(\tau,\gamma;\phi^*,\varphi^*) = \int_T{ e^{-j2\pi \gamma t}\mathbf{y}''(t;\phi^*,\varphi^*)\mathbf{U}^*(t-\tau)}dt,
\end{eqnarray} 
where $T$ is the signal duration of $\mathbf{y}''(t;\phi^*,\varphi^*)$. A peak appears when $\mathbf{U}(t-\tau)$ and $\mathbf{y}''(t;\phi^*,\varphi^*)$ align with each other in both time and frequency.

\parabreak \systemname combines the above individual modules into a four\hyp{}dimensional estimator shown in Fig.~\ref{f:MF_design}, for a two\hyp{}antenna transmitter and receiver.  We first estimate AoA at each timeslot,  before feeding the output of the AoA beamformer to the AoD estimator. The output of the AoD estimator is then passed to the Doppler estimator to estimate and remove any Doppler shift, and then correlated with a delayed transmit signal $\mathbf{U}(t-\tau)$ to estimate the ToF. The careful reader will notice that \systemname's processing in this part of the design is sequential.  However, the order of processing is carefully chosen: Doppler shift affects all antennas equally, and so the AoA and AoD estimators, which rely on measuring phase differences across different antennas, are unimpaired by arbitrary Doppler shifts.  Furthermore, the search in parameter space described in the next section allows us to jointly estimate all parameters.  For the general four\hyp{}dimensional estimator given in
Fig.~\ref{f:MF_design}, its output is: % given by:
\begin{eqnarray}
\label{eqn:z_function}
z(\phi,\varphi,\tau,\gamma) = \int_T{ e^{-j2\pi \gamma t} 
	\mathcal{F}^{-1}\{\mathbf{g}^H(\varphi)  \mathbf{H} \mathbf{c}(\phi)^*\odot \mathbf{LTF}\} \mathbf{U}^*(t-\tau)}dt ,\nonumber
\end{eqnarray}
where the $z(\phi,\varphi,\tau,\gamma)$, denoted as the \textit{z-function}, is calculated for all possible
$[\phi,\varphi,\tau,\gamma]$ combinations. The estimation of
parameters $[\phi,\varphi,\tau,\gamma]$  is then obtained by:
\begin{eqnarray}
\label{eqn:ML_theta}
(\phi,\varphi,\tau,\gamma)_{est} = \argmax_\upsilon 
{\left|z(\phi,\varphi,\tau,\gamma)\right|}.
\end{eqnarray}
Combining 
Eq.~\ref{eqn:s_signal} and Eq.~\ref{eqn:ML_theta} with
Eq.~\ref{eqn:noisySignal} then yields $\alpha$:
\begin{eqnarray}
\label{eqn:ML_alpha}
(\alpha)_{est} = \frac{1}{M\cdot T\cdot P} z((\phi,\varphi,\tau,\gamma)_{est}),
\end{eqnarray}
where $M$ is the antenna number, $T$ is the signal duration and $P$ is the transmit power. The received
signal $s(t)$ is now fully characterized by the parameter vector
$\upsilon=[\phi,\varphi,\tau,\gamma,\alpha]^T$.

Our design is module-based with a high flexibility. It can be adapted to a three\hyp{}dimensional $[\phi,\tau,\gamma]$ estimator for the single\hyp{}antenna transmitter case, and one\hyp{}dimensional ToF estimator for the single antenna transceiver case.

\subsection{Resolving multiple paths}
\label{s:resovMultipath}

The discussion in the preceding section has considered a single wireless propagation path. We now extend our design to handle multipath propagation.  We assume the antenna array receives signals from $L$ distinct paths, denoting the signal from the $l^{\mathrm{th}}$ path as $s(t;\upsilon_l)$ where $\upsilon_l = [\phi_l,\varphi, \tau_l,\gamma_l,\alpha_l]^T$. The received signal is thus the superposition of the signals from all $L$ paths:
\begin{eqnarray}
\label{eqn:recvSignal}
\mathbf{Y}(t) = \sum_{l=1}^L\mathbf{s}_l(t;\upsilon_l) + \mathbf{W}(t).
\end{eqnarray}
The goal is to estimate the path parameters: 
\begin{eqnarray}
\label{eqn:thetaAll}
\mathbf{V} = [\upsilon_1,\upsilon_2,\dots,\upsilon_L],  	
\end{eqnarray}
for all $L$ paths in $\mathbf{Y}(t)$. Note that the number of multipaths $L$ is also unknown and thus must be estimated.  

\begin{figure}
	\includegraphics[width=0.99\linewidth]{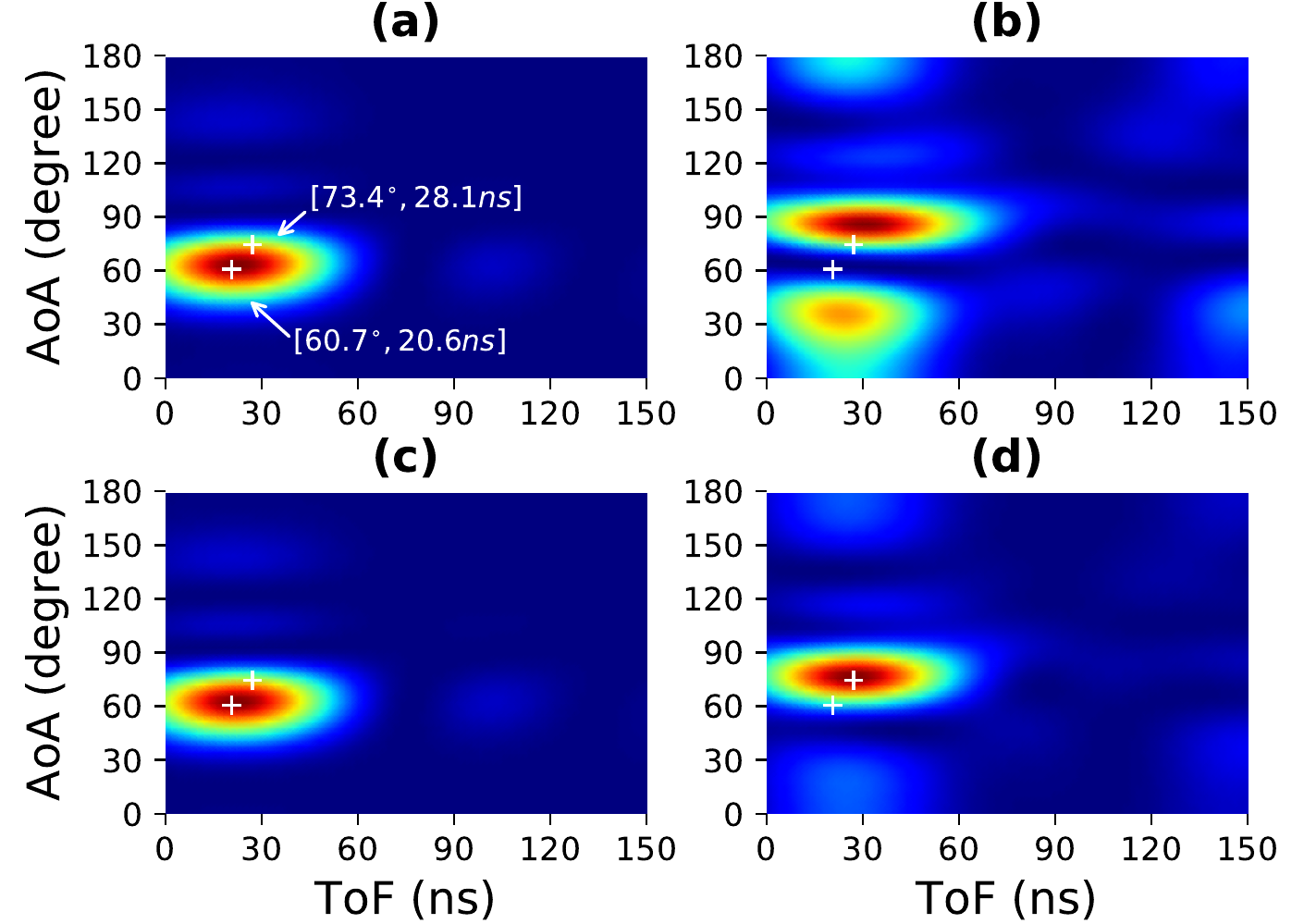}
	\caption{Output~(z-function) of the multi\hyp{}dimensional estimator with input: (\textbf{a}) superimposed signal of a stronger signal with AoA and ToF $[60.7^\circ,20.8~ns]$ and a weaker signal with $[73.4^\circ,28.1~ns]$; (\textbf{b}) the residual signal after canceling the stronger signal; (\textbf{c}) reconstructed stronger signal and (\textbf{d}) reconstructed weaker signal after three iterations. The peaks of z-function appear at (\textbf{a})$[63.5^\circ,22.5~ns]$, (\textbf{b}) $[85.9^\circ,30.5~ns]$, (\textbf{c}) $[61.2^\circ,21.3~ns]$, and (\textbf{d}) $[76.7^\circ,27.0~ns]$.}
	\label{f:zFun_2path}
	\vspace{-0.5cm}
\end{figure}

Resolvability is the premise of accurate path estimation. Non-resolvable signals are merged and the estimated parameters of the combined signal lead to a non-existing path which deviates from the true ones~\cite{tonetrack-mobicom15,ArrayTrack}. Although our multi\hyp{}dimensional estimator in Section~\S\ref{s:design:mde} improves the resolvability by higher dimensional parameters $[\phi_l,\varphi, \tau_l,\gamma_l,\alpha_l]^T$ of each path, the energies of nearby signal paths may still affect the estimation accuracy, e.g., a weaker signal may be overwhelmed by a nearby strong signal and thus not detected. We perform a controlled experiment (with the settings in Section~\S\ref{eval:resovlability}) to demonstrate such an effect, where
two signals with AoAs ($60.7^{\circ}$ and $73.4^{\circ})$ and ToFs (20.6 and 28.1~$ns$) are created. The signal with ToF $20.6~ns$ is 10~dB stronger than the other. We input the superimposed signal into our 2\hyp{}dimensional estimator (AoA and ToF) and obtain the z-function as depicted in Fig.~\ref{f:zFun_2path}~(a). It is expected that two peaks will appear in Fig.~\ref{f:zFun_2path}~(a), but the weaker signal, in practice, is dominated by the stronger signal and cannot be detected from the z-function. Similar results are obtained in previous studies using 1\hyp{}dimensional MUSIC or 2\hyp{}dimensional SpotFi~\cite{SpotFi}. A main reason is that those approaches identify the signal paths and estimate the parameters in a single round. No iterative refining  is included to accurately determine energy shares between different signal paths.

\systemname employs iterative parameter refinement during which the path signals are iteratively re-estimated and more accurately reconstructed from refined parameters in multiple rounds. We borrow the idea of successive interference cancellation (SIC) technique from data communications \cite{SIC1,SIC2} to stepwise estimate and subtract the signal of each path from the received signal and derive the initial estimates as a starting point of the refinement process.

\subsubsection{Initial estimation} 
\label{s:design:mp:initial}

To process the $L$ superimposed signals $[s_1(t)$, $s_2(t), \ldots$, $s_L(t)]$ in decreasing order of signal strength, we first treat all signals but the strongest, \ie, $s_1(t)$, as noise, applying the multi\hyp{}dimensional estimator to the raw received signal $Y(t)$ to obtain the output z-function, just as in Fig.~\ref{f:zFun_2path}(a). We estimate the parameters $\upsilon_1$ of signal $s_1(t)$ as the highest peak in the z-function. After this estimate, we reconstruct signal $s_1(t)$ using $\upsilon_1$ and cancel it from $Y(t)$.  We then calculate the \emph{residual} signal as: 
\begin{align}
\label{eqn:yt_all_min_1}
\mathbf{y}_2(t) = \mathbf{Y}(t) - \mathbf{s}_1(t) = \mathbf{s}_2(t) + \dots + \mathbf{s}_L(t)+ \mathbf{W}(t).
\end{align}
In this residual signal, the second strongest signal $\mathbf{s}_2(t)$ dominates. We then iterate the process, passing $\mathbf{y}_2(t)$ to the multi\hyp{}dimensional estimator, which now treats $\left[\mathbf{s}_3(t)\right.$, $\dots$, $\left.\mathbf{s}_L(t)\right]$ as noise. Now $\mathbf{s}_2(t)$ results in the highest peak in z-function, as shown in Fig.~\ref{f:zFun_2path}~(b). We continue to iterate until all the signals are separately estimated, stopping when the residual power in $\mathbf{y}_L(t)$ is below the dynamic range of the radio. The number of paths $L$ is also obtained. We note that, the final residual signal: 
\begin{eqnarray}
\label{eqn:noise_est}
\widehat{\mathbf{W}}(t) = \mathbf{Y}(t) - \sum_{l=1}^L \mathbf{s}_l(t),
\end{eqnarray}
is our estimation of the background noise.

\subsubsection{Iterative path parameter refinement}
\label{s:design:mp:refinement}

At this point, \systemname has obtained an estimate of the number of signal paths $L$ and a coarse estimate of each path's parameters. Remaining error arises from two sources. First, although small, interference from weaker signals still affects the estimation of stronger signals. As shown in Fig.~\ref{f:zFun_2path}(a), the peak of the z-function, \ie, $[63.5^\circ,22.5ns]$, is the estimation result for stronger signal, which deviates slightly from the ground truth, \ie, $[60.7^\circ,20.8ns]$, due to the existence of the nearby weaker signal. Second, the energy leakage from stronger signal during the cancellation introduces errors to the estimation of the weaker signal. The stronger signal we have reconstructed and canceled is therefore not identical to the true signal due to estimation errors. Energy leaks in the residual signal have a large effect on the estimation of weaker signal and can result in a large estimation error, as shown in Fig.~\ref{f:zFun_2path}(b).
\begin{figure*}
	\centering
	\includegraphics[width=0.92\linewidth]{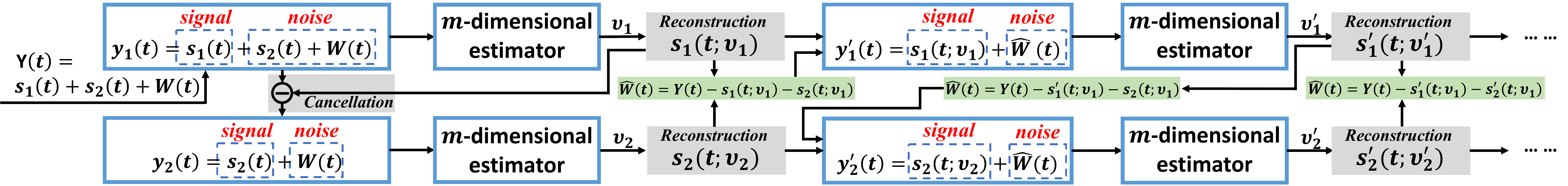}
	\caption{The framework of the iterative path parameter estimation algorithm ($L=2$ wireless propagation paths). SIC gives the coarse estimation $\upsilon_1$ and $\upsilon_2$. The iterative process keep refining the estimation till convergence.}
	\label{f:SIC}
%	\vspace{-0.5em}	
\end{figure*}
%In order to refine the path parameter estimates, we pass the received signal through an iterative process where we alternatively re\hyp{}estimate the multi\hyp{}dimensional path parameters and reconstruct individual path signals from those estimates, as we now describe.
To remove the interference, we reconstruct the stronger signal $\mathbf{y}_l(t)$ from $l^{\mathrm{th}}$ path with its estimated parameters and the estimated noise as:
\begin{eqnarray}
\label{eqn:xt_est}
\mathbf{y}_l^\prime(t) = \mathbf{s}_l(t;\upsilon_l) +  \widehat{\mathbf{W}}(t),
\end{eqnarray}
where the $\mathbf{s}_l(t;\upsilon_l)$ is reconstructed using the parameters $\upsilon_l$ obtained in the first round. %  with our interference cancellation, according to Eq.~\ref{eqn:s_signal}. 
The signal $\mathbf{y}^\prime_l(t)$ is now refined, containing less interference compared to the raw signal $\mathbf{y}_l(t)$ in which weaker signals are also considered as noise. We then re-estimate the path parameters of $\mathbf{y}'_l(t)$, $\upsilon'_l$, with the multi\hyp{}dimensional estimator given in \S\ref{s:multiEst}. To reduce the leakage, we keep calculating and updating the noise. Supposing we have obtained new parameters $[\upsilon'_1,\upsilon'_2,...,\upsilon'_{l_s}]$ for $l_s$ stronger paths, we can update the noise of the ${(l_s+1)}^{\mathrm{th}}$ path as:
\begin{eqnarray}
\label{eqn:noise_est_update}
\widehat{\mathbf{W}}(t) = \mathbf{Y}(t) - \sum_{l=1}^{l_s} \mathbf{s}^\prime_l(t;\upsilon_l^\prime) - \sum_{l=l_s+1}^{L} \mathbf{s}_l(t;\upsilon_l),
\end{eqnarray}
where the $\mathbf{s}_l^\prime(t;\upsilon_l^\prime)$ is reconstructed using refined parameters $\upsilon_l^\prime$ and is closer to the true received signal so that less leakage is contained in the noise. The signal of ${(l_s+1)}^{\mathrm{th}}$ path is then reconstructed using updated noise according to Eq.~\ref{eqn:xt_est}.

The path re\hyp{}estimation and reconstruction is performed for every path, and so we obtain the estimate $\mathbf{V}' = \left[\upsilon'_1,\dots,\upsilon'_L\right]$ for all paths. Now we have obtained a better estimate $V'$, we feed it back into Eq.~\ref{eqn:xt_est} to start another round of estimation. The above steps of multipath signal re\hyp{}estimation and reconstruction are iterated, and stop when the difference of each path parameter between two consecutive iterations is smaller than a predefined threshold. The threshold can be flexibly tuned to meet different levels of accuracy requirements. In our current implementation, the threshold is set as the step size of the parameter search in \S\ref{s:computationOverhead}, yielding accurate results for passive localization and motion tracking (\S\ref{s:eval}).

Fig.~\ref{f:SIC} illustrates the workflow of the iterative process based on an example of superimposed signal from two paths. The initial signal cancellation and estimation steps obtain the coarse estimates $\upsilon_1$ and $\upsilon_2$ for the two paths. From there, the iterative process reconstructs the path signals $\mathbf{y}'_1(t)$ and $\mathbf{y}'_2(t)$, re-estimates the path parameters $\upsilon_1^{\prime}$ and $\upsilon_2^{\prime}$, and updates the residual signal $\widehat{\mathbf{W}}(t)$ in each iteration. Fig.~\ref{f:zFun_2path}s~(c) and (d) present the z-function of our multi\hyp{}dimensional estimator after three rounds of iteration. It is clear that the peak in Fig.~\ref{f:zFun_2path}s~(c) and (d) is much closer to the true path parameters compared with Fig.~\ref{f:zFun_2path}~(a) and (b).

\paragraph{Convergence.} \label{para:convergence} A common issue for iterative algorithms is their convergence. It can be proved\footnote{The proof is not included due to page limitation.} that the above iterative process belongs to the EM family of algorithms~\cite{EM,EM1} with the expectation step given by Eq.~\ref{eqn:xt_est} and maximization step given by Eq.~\ref{eqn:ML_theta}, and so convergence is guaranteed~\cite{EM1}. One problem with EM is that it may converge to a local instead of the global maximum if the EM algorithm is not properly initialized.  Comprehensive experiments in \S\ref{s:eval} empirically show that our cancellation\hyp{}based initialization is able to provide an accurate initial start, ensuring that our iterative algorithm almost always converges to the global maximum.

\subsection{Reducing computational complexity}
\label{s:computationOverhead}
%\paragraph{Compuation analysis.} 
From Fig.~\ref{f:SIC}, we see that the multi\hyp{}dimensional estimator accounts for a significant portion of \systemname's computation complexity. Each multi\hyp{}dimensional estimator solves am ML problem by an exhaustive search. Specifically, the total number of possible combinations that a 4\hyp{}dimensional estimator that estimates $[\phi,\varphi,\tau,\gamma]$ has to access is given by:
\begin{eqnarray}
	\label{eqn:overhead}
	\eta =  p_\phi \times p_\varphi \times p_\tau \times  p_\gamma,
\end{eqnarray}
where $p_\phi$, $p_\varphi$, $p_\tau$ and $p_\gamma$ are the number of steps for each path parameter, respectively (\eg, $p_\phi = 100$ if the range for AoA estimation is $[0,\pi]$ with a step size of $0.01\pi$). The number of combination $\eta$ increases exponentially with the number of dimensions, causing unbearable overhead. From Fig.~\ref{f:SIC}, we see that the estimator is applied $N_{\mathrm{iter}}\times L$ times, supposing we have $L$ paths and the system iterates $N_{\mathrm{iter}}$ rounds.

During each iteration, the input to each multi\hyp{}dimensional estimator is actually the signal from one path plus the noise, as shown in Fig.~\ref{f:SIC}. Normally, the power level of the signal is tens of dB higher than noise, so we observe a dominant peak in the output z-function of the estimator as Figures~\ref{f:zFun_2path}~(c) and (d) show. Based on this observation, we use a \emph{coordinate descent} method~\cite{coordinateDescent} to approach the maximum. We replace the four\hyp{}dimensional search in Eq.~\ref{eqn:ML_theta} with four one\hyp{}dimensional searches. Specifically, we fix three of the four parameters (\eg $\varphi$, $\tau$, and $\gamma$) and search for the value of the fourth parameter(\eg, $\phi$) that can generate a maximal output:
\begin{align}
\label{eqn:ML_phi}
\phi^{\prime\prime} &= \argmax_\phi {\left|z(\phi,\varphi^\prime,\tau^\prime,\gamma^\prime)\right|},
\end{align}
where $\varphi^\prime$, $\tau^\prime$ and $\gamma^\prime$ are the estimates from the previous iteration. We repeat this process for the other three parameters:
\begin{align}
\label{eqn:ML_phi_gamma}
\varphi^{\prime\prime}   &= \argmax_\varphi {\left|z(\phi^{\prime\prime},\varphi,\tau^\prime,\gamma^\prime)\right|}, \\
\tau^{\prime\prime}      &= \argmax_\tau {\left|z(\phi^{\prime\prime},\varphi^{\prime\prime},\tau,\gamma^\prime)\right|}, \\
\gamma^{\prime\prime}    &= \argmax_\gamma{\left| z(\phi^{\prime\prime},\varphi^{\prime\prime},\tau^{\prime\prime},\gamma)\right|},
\end{align}
and obtain an update of all four parameters. By doing so, we reduce the search space from $p_\phi \times p_\varphi \times p_\tau \times p_\gamma$ to $p_\phi + p_\varphi + p_\tau + p_\gamma$, a significant reduction.  The trade-off is that the global maximum of each multi-dimensional estimator in Fig.~\ref{f:SIC} can not be guaranteed~\cite{GEM,GEM1}. We can only guarantee an increase in the output z-function compared with the start point instead of maximizing it.
 
\textit{Convergence of \systemname.} From Fig.~\ref{f:SIC} we see that $L$ multi\hyp{}dimensional estimators are needed in each round of iterations. If every multi\hyp{}dimensional estimator in this round maximizes its z-function, then the expectation is maximized and \systemname walks a big step towards the optimal. We relax the requirement of maximization to simply increasing the expectation, which is called generalized EM~(GEM)~\cite{GEM,GEM1}. Coordinate descent method satisfies the increasing of z-function and \systemname becomes a GEM.  For a GEM algorithm, the convergence to maximum (either local or global maximum) is still guaranteed but more iterations may be needed to achieve the final maximization of expectation~\cite{GEM,GEM1}. Nevertheless, since we carefully initialize the algorithm and select a reasonably good starting point with the initial cancellation and estimation steps, \systemname still converges fast. Our experimental results
show that \systemname converges in less than 9 rounds in 90\% of cases, even with 10 signal paths. The number of iterations typically reduces to as small as five when there are five signal paths. Therefore, the overall computational overhead is significantly reduced.

 \section{Channel Measurement}
\label{s:chMea}

In this section, we introduce how we estimate the wireless channel
using Wi-Fi transceivers. Channel measurement errors are inevitable
due to imperfect hardware, and so we also describe how we handle channel
measurement uncertainty.

In order to estimate one channel parameter, we have to sample the
wireless channel in its corresponding sampling domain. For example, to
estimate ToF, we must sample the channel in the frequency domain.
Similarly, to estimate a frequency shift that has been added to
the signal (\eg, a Doppler frequency shift), we need to sample the
channel in the time domain. Channel sampling in the spatial domain is
required to estimate signal angle (including AoA and AoD).
Table~\ref{t:CH_sampling} summarizes the channel parameter and its
corresponding sampling domain. 
\begin{table}[h]
	\centering
	\begin{tabularx}{\linewidth}{@{}Xlll@{}}
		\toprule
{\bf Channel	parameter:} &$\phi,\varphi$& $\tau$& $\gamma$ \\ 
{\bf Sampling domain:} & Space & Frequency & Time\\  
		\bottomrule
	\end{tabularx}
	\caption{Channel parameters and their
  corresponding channel sampling domains.}
		\label{t:CH_sampling}
		\vspace{-1em}
\end{table}

To estimate multiple parameters, we need to sample the channel in multiple domains. Fig.~\ref{f:SampleRet} depicts the channel sampling results for estimating AoA + AoD + ToF + frequency shift. Where $N$ and $M$ are the numbers of transmitting/receiving antennas (spatial domain), $F$ is the subcarrier number (frequency domain), and $T$ is the number of channel samples in time (time domain). If the interval between two channel sampling is $t_s$, then we sample the channel for the duration of $T\cdot t_s$. The channel sampling results is a matrix of size $N\times M \times F \times T $. Any sub-matrix can be used to estimate a subset of parameters. For example, a channel sampling matrix of size $M\times F$ can be used to estimate AoA + ToF. We note that our parameter estimation algorithm, described in the previous section, can take any dimensional matrix as input and estimate any combination of parameters.

\subsection{Measurement Error Handling}
\label{s:ChanM_error}
Since the hardware we use to measure the wireless channel is imperfect,
channel measurement errors are introduced inevitably. For example, it has
been observed that a constant, time-invariant phase offsets exist for the signals transmitted or received at sender and receiver~\cite{ArrayTrack},
which affect estimates of AoD and AoA, respectively. Sampling
frequency offset (SFO) and symbol timing offset (STO) affect
the estimation of ToF. Carrier frequency offset (CFO) itself is a
frequency shift that added to the signal, which appears as an actual Doppler
frequency shift. Technically speaking, CFO is not an error but will
affect the estimation of Doppler shift without proper handling.
Table~\ref{t:CH_sampling_err} summarizes error sources and the
estimation of each parameter they affect. 

\begin{table}[h]
	\centering
	\scriptsize
	\begin{tabular}{ M{1.6cm} M{1.32cm} M{1.3cm} M{1.2cm} M{1.0cm} }
		\toprule
	{\bf Error source:}      & Phase offsets of TX-chains
  & Phase offsets of RX-chains                 & SFO, STO  & CFO\\ 
	{\bf Parameter:} &  AoD              & AoA        & ToF           & Doppler\\  
		\bottomrule
	\end{tabular}
	\caption{Channel measurement error source and the corresponding channel parameter it affects.}
	\label{t:CH_sampling_err}
	%\vspace{-0.2em}
	\vspace{-1.2em}
\end{table}

\paragraph{Phase offset across radio chains.} 
The wireless signal transmitted or received by multiple antennas experiences a phase shift introduced by the radio chains~\cite{ArrayTrack}. The phase offsets across radio chains must be eliminated to provide accurate angle estimation (AoA and AoD). Such a phase offset is constant across time ($T$ samples) and frequency ($F$ subcarriers), and can be measured by connecting the transmit chain and receive chain of WARP or COTS Wi-Fi devices via a coaxial cable, as Fig.~\ref{f:phaseCalib}(a) illustrates, where three transmit chains and receive chains are connected. We, however, observe that spatial multiplexing fails and no packets can be correctly received with such a setup as the channel matrix becomes singular. To break the singularity, we use the setups shown in Fig.~\ref{f:phaseCalib}~(b) and (c). In Fig.~\ref{f:phaseCalib}~(b), we measure the phase offset between: 1) transmit chain 2 and transmit chain 3, denoted as $\alpha_2$; 2) receive chain 1 and receive chain 2, denoted as $\beta_1$. Similarly, phase offsets $\alpha_1$ and $\beta_2$ are measured with the setups shown in Fig.~\ref{f:phaseCalib}(c). The phase offsets are then canceled with the measured value.
\begin{figure}[t]%[htb]
	\centering
	\includegraphics[width=0.85\linewidth]{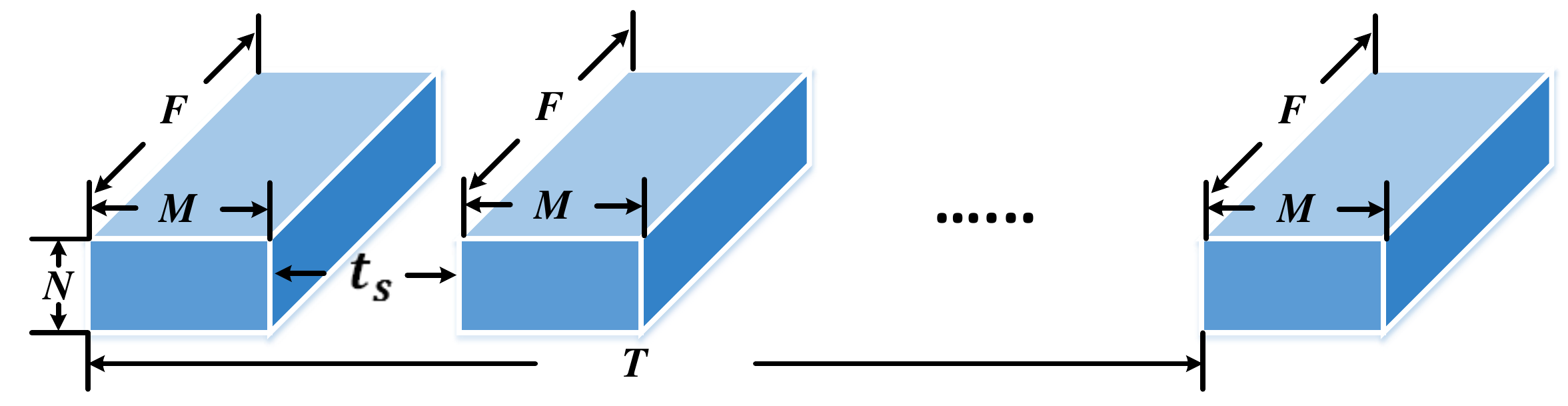}
	\caption{Channel sampling results for jointly estimating four channel parameters (AoA + AoD + ToF + frequency shift).}
	\label{f:SampleRet}
% 	\vspace{-0.5em}
\end{figure}
\begin{figure}[h]%[htb]
	\centering
	\includegraphics[width=0.9\linewidth]{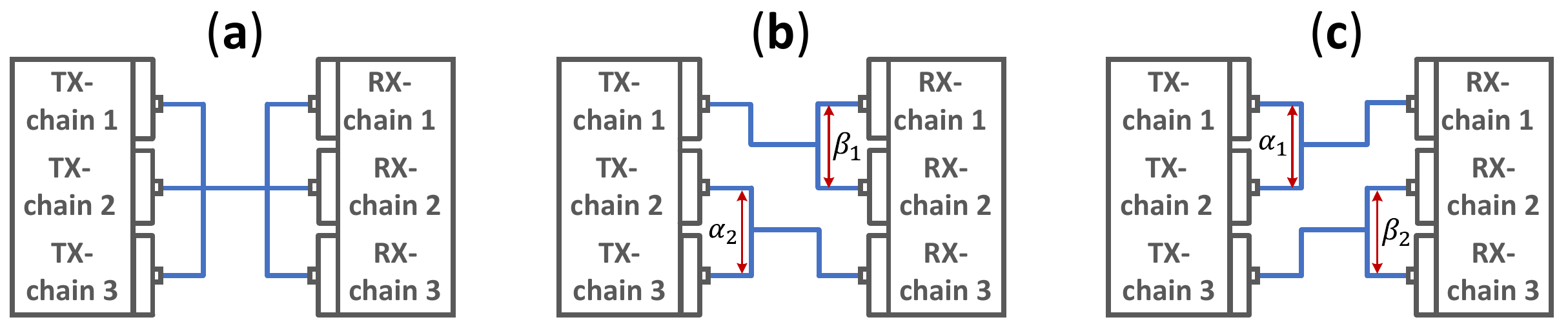}
	\caption{Three TX chains and RX chains are connected together with power splitter/combiner in (\textbf{a}). One TX or RX radio chains are connected to two TX or RX radio chains in (\textbf{b}) and (\textbf{c}).
% 	in (\textbf{a}); One tx-chain is connected with two rx-chains and the rest two tx-chains are connected with the another rx-chains in (\textbf{b}) and (\textbf{c}).
}
	\label{f:phaseCalib}
	%\vspace{-0.5em}
\end{figure}

\paragraph{SFO and STO} Due to the lack of tight time synchronization between Wi-Fi transceivers, SFO and STO introduce phase errors in the frequency domain and thus affect the estimation of ToF~\cite{tonetrack-mobicom15,Yaxiong-mobicom15}. The SFO and STO are the same across antennas of the same channel sample (one packet) but vary across samples, which results in different phase errors added to the $T$ channel samples of Fig.~\ref{f:SampleRet}. Estimation based on such a matrix won't give us meaningful ToF results. We thus propose to use the phase of the first channel sample as an anchor and then align the phase of the rest $T-1$ samples to it. Specifically, the phase error introduced by SFO and STO can be modeled as $e_k = k\cdot \lambda$, where $k$ is the index of the subcarrier. The error is added linearly to the phase: $P_k = \theta_k + e_k$, where $P_k$ and $\theta_k$ are the measured and the real phase of subcarrier $k$, respectively. The exact value of slope $\lambda$ is decided by STO and SFO and is hard to measure. We thus calculate the slope difference between $i$th channel sample and the first channel sample by $P_{k,i}-P_{k,1}=k \cdot(\lambda_i - \lambda_1)$. We then remove the difference to align the phase errors in the frequency domain.

%and thus add a time\hyp{}varying delay to the estimated ToF value, which makes it practically difficult to obtain the absolute ToF~\cite{SpotFi}.
\begin{figure}[h]%[htb]
	\centering
	\includegraphics[width=0.97\linewidth]{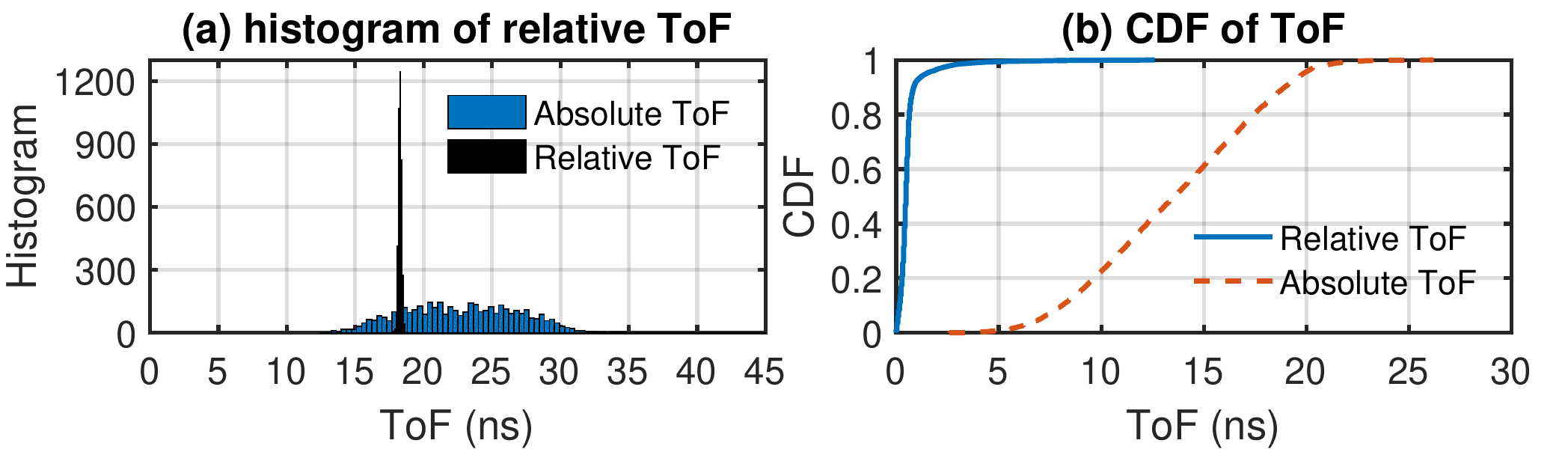}
	\caption{A microbenchmark study of two paths: \textbf{a}) the histogram of the absolute ToF and the relative ToF between two paths; (\textbf{b}) the CDF of the ToF error.}
	\label{f:relativeToF}
% 	\vspace{-1.5em}
\end{figure}
The phase error introduced by SFO and STO is aligned but not removed, which makes it practically difficult to obtain the absolute ToF~\cite{SpotFi}. We observe that the time delay added by SFO and STO is the same for all the paths. Therefore, the ToF difference between a pair of two paths is invariant even in the presence of SFO and STO. Fig. 7 shows the results from our controlled experiment by connecting radio chains of two QCA9558 together using two RF coaxial cables of different lengths to  generate different ToFs ($t_1=9.2~ns$ and $t_2=27.4~ns$). We then estimate the ToF of the received signal. Fig.~\ref{f:relativeToF}~(a) plots the histogram of the estimated absolute ToF values $t_1$ of the signal from the shorter cable, which exhibits a large variation across different transmissions. On the other hand, we see from this figure that the variation of the relative ToF $|t_2-t_1|$ between the two paths is minimal. Fig.~\ref{f:relativeToF}~(b) presents the CDF of the ToF estimation error. We see that even when the absolute ToF measurements are inaccurate (a median error of 13~ns), the relative ToF estimates can be very accurate (a median error of 0.48~ns).

 Based on the above observation, we use the ToF measurement of the direct path as a reference basis to calibrate the ToF estimations of other reflection paths. We first calculate the ground\hyp{}truth ToF, AoA, AoD of the direct path based on the locations of the transceiver pair. From all signal paths output from mD-Track, we identify the path with shortest ToF and largest amplitude, as the direct path and derive the $\Delta\tau$ between the measured and ground-truth ToF. We then use the derived $\Delta\tau$ to correct the ToF measurements for all reflection paths. In this way, we eliminate the impact of SFO and STO, and thus obtain accurate ToF information.

\paragraph{CFO} 
Both CFO and Doppler shift are frequency shifts added to the signal and so our estimator is unable differentiate the two. There are, however, two major differences between CFO and Doppler: 1) the magnitude, \ie, CFO is on the order of 100s of Hz and Doppler introduced by a human is only a few Hz; 2) CFO is added to all the multipaths, but Doppler is only introduced to mobile paths. Two important parameters determine the performance of frequency shift estimation. First, the channel sampling interval $t_s$ as Fig.~\ref{f:SampleRet} shows, which gives the maximum frequency shift $f_m$ that can be estimated by $f_m=1/t_s$. Second, the total sampling period $T\cdot t_s$, which gives the minimum frequency difference the estimation can differentiate by $1/(T\cdot t_s)$. The standard Wi-Fi protocol uses two consecutive LTF of one preamble to estimate the CFO, which has small sampling interval (and thus a large frequency range to estimate CFO of 100s of Hz) and small total sampling period (and thus coarse frequency resolution). We reuse the CFO estimation of standard Wi-Fi to get a coarse estimation of CFO and then remove it from the channel sampling results before the actual estimation so that it becomes much smaller. We then sample the channel with multiple packets, just as Fig.~\ref{f:SampleRet} shows, with a larger sampling interval $t_s = 25ms$ (smaller estimation range of 40Hz) and larger sampling period $1s$ (higher frequency resolution of 1Hz). With such a high frequency resolution, the small frequency shift introduced by human and the small residual CFO can be estimated. As CFO is constant for all paths, we subtract the frequency shift of the direct path (a static path) from the frequency estimation of all the paths. A residual frequency shift from Doppler effect remains.

 \section{Implementation}
\label{s:impl}
\begin{figure}[t]%[htb]
	\centering
	\includegraphics[width=0.99\linewidth]{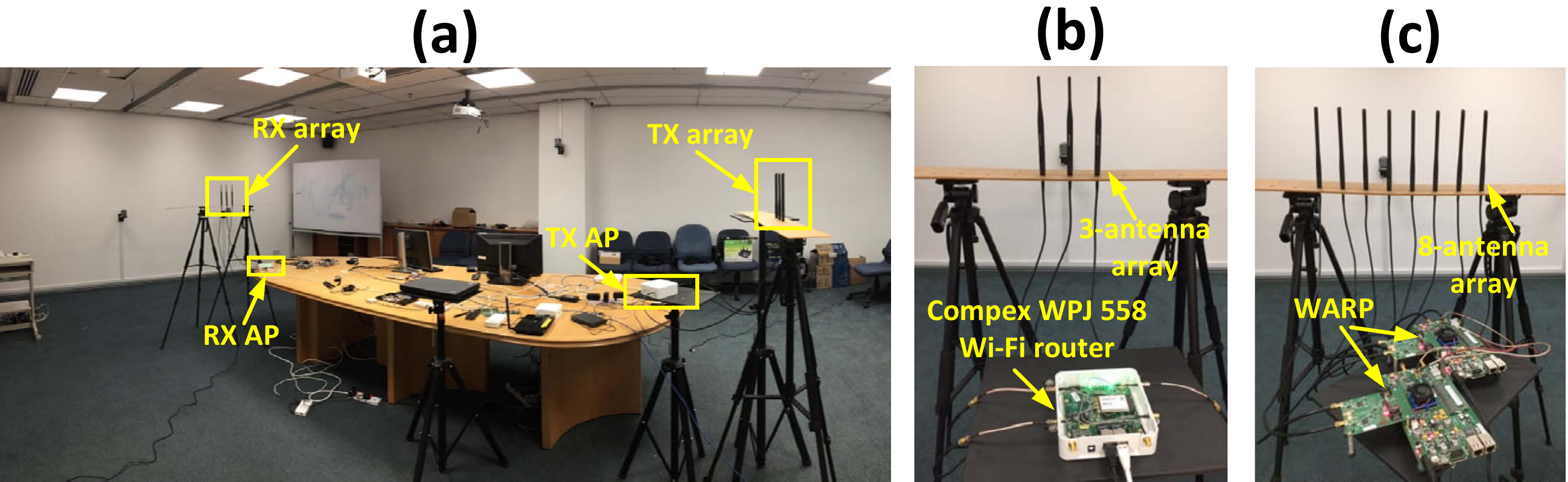}
	\caption{\textbf{(a)} Experimental set up in a indoor meeting room. We use both \textbf{(b)} COTS Wi-Fi AP with three antennas and \textbf{(c)} WARP with at most eight antennas as radio frontend to collect channel measurements. }
	\label{f:testbed}
	%    \vspace{-1.5em}
\end{figure}
\begin{figure*}[t]%[htb]
	\centering
	\includegraphics[width=0.91\linewidth]{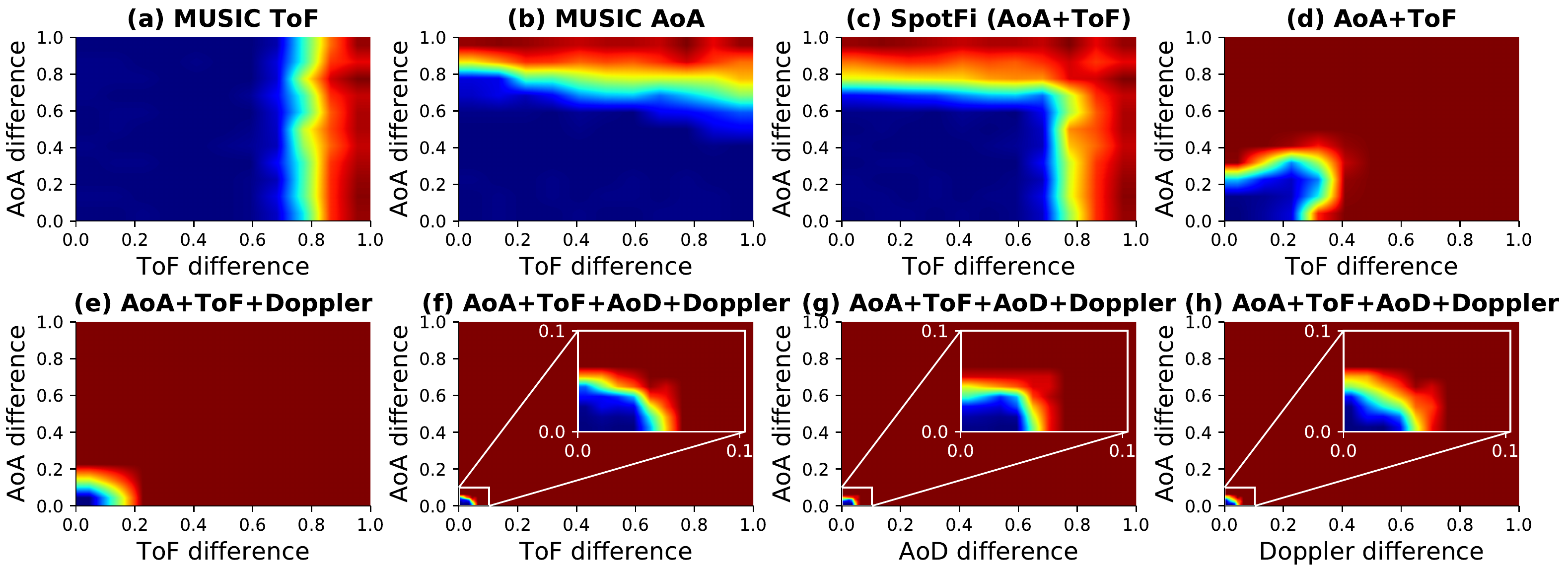}
	\caption{The color indicates the probability of resolving two signals: blue indicates ``non-resolvable" and red means ``fully resolvable". We compare the resolvability of: \textbf{(a)} MUSIC 1D to estimate ToF; \textbf{(b)} MUSIC 1D to estimate AoA; \textbf{(c)} SpotFi 2D to jointly estimate AoA+ToF; \textbf{(d)} mD-Track 2D to jointly estimate AoA+ToF; \textbf{(e)} mD-Track 3D to jointly estimate AoA+ToF+Doppler with Doppler unchanged; \textbf{(f)} mD-Track 4D to jointly estimate AoA+ToF+Doppler+AoD, with AoD and Doppler unchanged; \textbf{(g)} mD-Track 4D, with ToF and Doppler unchanged; and \textbf{(h)} mD-Track 4D with ToF and AoD unchanged. }
	\label{f:ResolutionLimit}
	\vspace{-0.4cm}
\end{figure*}
We implement the core estimation algorithms of \systemname at a backend server which is a desktop workstation. We use both WARP v3 boards~\cite{warp} and commodity Wi-Fi routers to sample the channel and send the results back to server.

\noindent\textbf{WARPs.} WARP v3 boards collect time domain IQ samples and send them to a PC connected to the WARP via an Ethernet cable. The PC detects the begining of the packet, extracts the LTF and sends the LTF to the server.

\noindent\textbf{Commercial APs.} All of our APs are equipped with Atheros Wi-Fi NIC or SoC (AR9340, AR9580, and QCA9558), and run a customized OpenWRT system. We modify the Wi-Fi driver in kernel space of OpenWRT to enable CSI collection~\cite{Splicer}. We build a user-space application that retrieves the received frequency domain LTF by multiplying the transmitted LTF (defined in 802.11 standard) with received CSI (empty subcarriers of CSI and LTF are padded with zeros). We transform the frequency domain LTF into time domain sample via the IFFT, which are sent to the server over the backhaul network.

\section{Evaluation}
\label{s:eval}
 
We conduct experiments in different indoor environments including labs (600 and 420 $m^2$), meeting rooms (54 and 32 $m^2$) and corridors. Fig.~\ref{f:testbed}~(a) shows the experimental setup in one of our indoor meeting rooms. Channel measurements collected by commercial APs (Fig.~\ref{f:testbed}~(b)) and WARP (Fig.~\ref{f:testbed}~(c)) are sent to the server via wired backhaul, where parameters are estimated. All APs work in MIMO mode when using multiple antennas.
%The server estimates the path parameters, localizes the target and displays the localization results in real-time. The server also records the localization and parameter estimation results for further analysis.
\begin{table}[t]
	\centering
	\scriptsize
	\begin{tabular}{ P{0.81cm}  M{1.5cm}  M{1.5cm}    M{1.6cm}  M{1.2cm}}
		\toprule
		{\bf Version}   & {\bf mD-Track 2D}                  & {\bf mD-Track 3D }                &{\bf mD-Track 4D} & {\bf SpotFi}\\ 
%		\hline
		\toprule
		{\bf Parameter} & AoA + ToF              & AoA + ToF + Doppler         & AoA + ToF + AoD + Doppler & AoA + ToF\\  
		\hline
%		\toprule
		{\bf Input}     &$ M\times64$           & $M \times 64 \times T$             & $N \times M \times 64 \times T$  & $M \times 56$      \\    
		\bottomrule
	\end{tabular}
	\caption{We implement three different versions of \systemname as well as SpotFi for comparisons.}
	\label{t:Version}
	\vspace{-1cm}
\end{table}

\paragraph{Multiple versions of \systemname.} We implement three versions of \systemname, as shown in Table~\ref{t:Version}. \systemname 2D uses just AoA and ToF, like SpotFi, while \systemname 3D adds Doppler and \systemname 4D adds both Doppler and AoD. The input channel measurement to each version of \systemname is different. According to Fig.~\ref{f:SampleRet} and Table~\ref{t:CH_sampling}, a channel measurement matrix with size $M\times64$ (20~MHz) is used for \systemname 2D. The matrix size increases to $M\times64\times T$ and $M\times N\times64\times T$ for \systemname 3D and 4D respectively.

\paragraph{SpotFi.} We compare the passive localization performance with SpotFi, which jointly estimates the AoA plus ToF for localization\footnote{Wi-Deo~\cite{WiDeo} is another passive localization system using estimated path parameters. As we do not have the source codes of Wi-Deo, we do not directly compare with it.} Spatial and time smoothing is applied to SpotFi. We note that SpotFi is actually able to estimate the parameters of all the paths, including the reflection paths. We therefore make use of the parameters of the reflection path that SpotFi estimates to passively localize objects. To provide a fair comparison, we use the same input channel measurement matrix of $M\times 64$ for 2D SpotFi and 2D \systemname. Keeping increasing the matrix for SpotFi, \eg, $N\times 64 \times T$, only provides a marginal improvement for SpotFi since the time domain sampling works for estimation of frequency shift not the AoA or ToF, as Fig.~\ref{f:SampleRet} and Table~\ref{t:CH_sampling} shows.
  
\begin{figure*}[t]%[htb]
	\centering
	\includegraphics[width=0.95\linewidth]{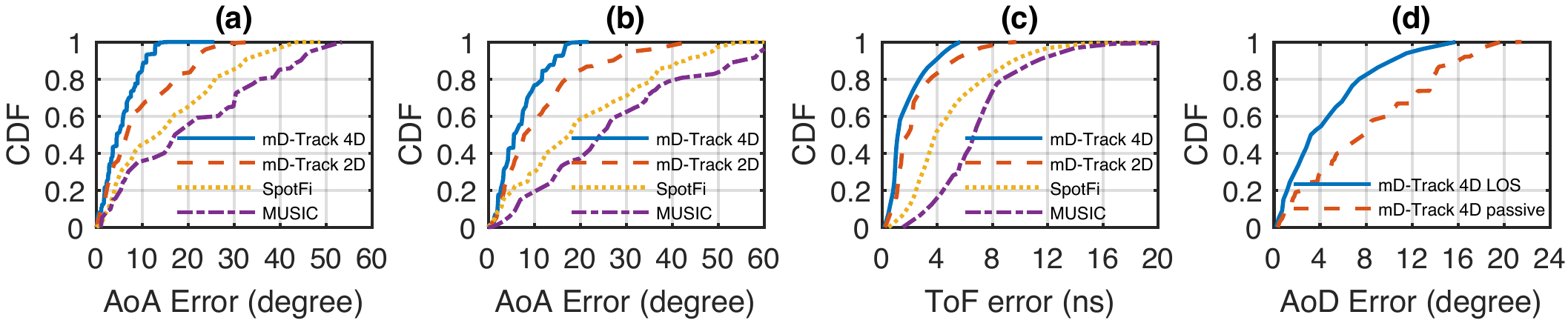}
	\caption{mD-Track runns on COTS AP. CDF of \textbf{(a)} AoA estimation errors of the direct path signal; \textbf{(b)} AoA estimation errors of reflection path signals; \textbf{(c)} ToF estimation errors; \textbf{(d)} AoD estimation errors, for mD-Track, SpotFi, and MUSIC. }
	\label{f:aoa_tof_dop_cdf}
	\vspace{-0.5em}
\end{figure*}

\subsection{Resolving multipaths}
In this section, we first present a resolvability analysis of \systemname which studies its capability to resolve two paths that has similar parameters. We then study the estimation accuracy of the path parameters in each dimension.

\subsubsection{Resolvability}
\label{eval:resovlability}
 
We demonstrate that \systemname has a significantly better ability than previous algorithms to resolve two signals.  We conduct controlled experiments by connecting the
transmitting and receiving RF chains of two WARPs with coaxial cables, varying the length of cable to provide different propagation times. We rotate the transmitted signal phases to emulate different AoAs, AoDs and Doppler shifts. We vary the ToF, AoA, AoD and Doppler differences between the two signals by a multiple of the parameter's \emph{basic resolution}: the resolution of ToF at a 20~MHz bandwidth is $\Delta\tau=50~ns$, and Doppler at a $1~s$ observation time is $\Delta\gamma=1$~Hz. Angular resolution of AoA and AoD is
$\Delta\phi=14.2^\circ$ for eight antennas in the array~\cite{AngularResolution}. When two paths are close to each other, they may not be able to be resolved and may be estimated as one path deviated from both the two paths. 
For each parameter configuration, we transmit 1,000 packets and resolve the two signals with \systemname 1,000 times using received packets.

Fig.~\ref{f:ResolutionLimit} plots the results. The x-axis and y-axis are the path parameter differences relative to basic resolution in that dimension, \eg, 0.2 of ToF means $0.2\times50~ns=10~ns$ and the color depth (from blue to red) indicates the probability that two path signals are resolvable in 1,000 estimations. We compare the resolvability of \systemname with SpotFi and MUSIC. %Our observations from Fig.~\ref{f:ResolutionLimit} are as follows. 
We observe that all algorithms perform better than the basic resolution limit so that two signals with parameter difference smaller than its basic resolution can still be resolved.
Furthermore, increasing the number of dimensions significantly increases resolvability.  One dimensional MUSIC is not able to resolve two signals with AoA\fshyp{}ToF difference smaller than  0.75 of its basic resolution. The fraction can be reduced to about 0.4 if applying 2D~\systemname, 0.1 if applying 3D~\systemname and 0.04 if applying 4D~\systemname. At last, even with the same dimension, 2D \systemname still outperforms 2D~SpotFi in Fig.~\ref{f:ResolutionLimit}~(c) and (d). The improvement stems from the iterative interference cancellation and path refinement process in \systemname. %Both MUSIC and SpotFi estimate parameters of all resolvable paths simultaneously and ignore the mutual interference between paths. 
\textbf{Takeaway.} \textit{Increasing the dimension improves the resolvability significantly.}

\subsubsection{Estimation Accuracy}

We demonstrate that \systemname can estimate path parameters at high accuracies even using commodity Wi-Fi APs with only three antennas in this section. We collect channel readings using Compex WPJ558 router. We calculate the ground truth of AoA, AoD, and ToF based on the actual positions of the transceivers. We compare AoA and ToF estimation with MUSIC and SpotFi and favor MUSIC and SpotFi by selecting the path with the AoA estimate closest to the ground truth as their estimate. 

\parahead{AoA and ToF estimation} AoA estimation results are presented in Figs.~\ref{f:aoa_tof_dop_cdf}~(a) and (b). \systemname provides significantly more accurate AoA estimates: with only three antennas, \systemname's median AoA error for the direct path is as small as $4.4^\circ$ and $6.2^\circ$ on 4D and 2D versions respectively, compared to $13.4^\circ$ and $17.1^\circ$ from SpotFi and MUSIC. The AoA estimates of the weaker reflection paths are less accurate. Despite that, \systemname still achieves a median AoA error of $5.6^\circ$ and $7.3^\circ$ on 4D and 2D versions respectively, compared with $16.9^\circ$ and $24.2^\circ$ for SpotFi and MUSIC. 
From Fig.~\ref{f:aoa_tof_dop_cdf}~(c), we see that \systemname 4D and 2D are able to estimate ToF with a median error of $1.23~ns$ and $1.85~ns$, while SpotFi and MUSIC achieve a median error of $3.8~ns$ and $6.7~ns$.

\parahead{AoD estimation} Fig.~\ref{f:aoa_tof_dop_cdf}(d) gives \systemname 4D's AoD estimation results for LoS path and reflection path. The median errors are $3.3^\circ$ and $7.6^\circ$ respectively. According to our knowledge, \systemname is the first system based on COTS Wi-Fi hardware that can estimate AoD at such a high accuracy. 

\begin{figure}[t]%[htb]
	\centering
	\includegraphics[width=0.93\linewidth]{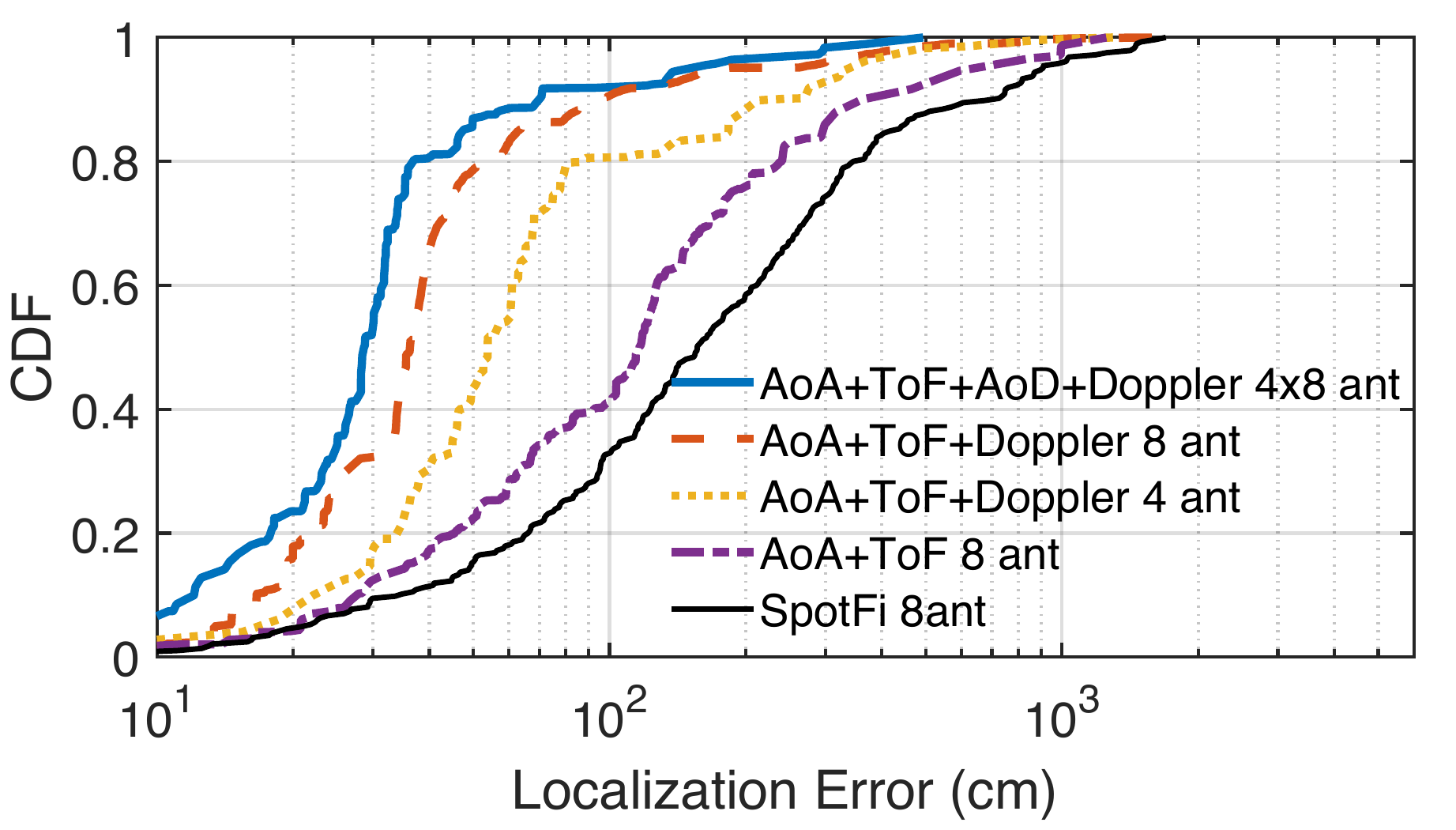}
	\caption{Passive localization error of \systemname and SpotFi when using the WARP and a 40~MHz channel.}
	\label{f:WARPlocErr}
	 \vspace{-1.1em}
\end{figure}

\subsection{Passive Localization}
In this section, we evaluate the device-free passive localization performance of \systemname with a single pair of wireless terminals (WARP or COTS Wi-Fi APs).

\parahead{WARP performance} The results obtained using WARP are presented in Fig.~\ref{f:WARPlocErr}.  \systemname 3D and 4D are respectively able to achieve median localization errors of $0.36~m$ and $0.28~m$, with eight antennas and a 40~MHz bandwidth. To the best of our knowledge, no prior Wi-Fi based systems have demonstrated such a high accuracy with a single transceiver pair for passive localization.  We also compare \systemname with SpotFi. 
For a fair comparison, we implement an AoA and ToF joint estimator based on SpotFi's algorithm and modify it to work with reflected signals. The median error of SpotFi is $1.56~m$ with eight antennas and 40~MHz.  We can see that even \systemname 2D already achieves significantly better localization performance than SpotFi.
\begin{figure}[t]
	\centering
	\includegraphics[width=0.93\linewidth]{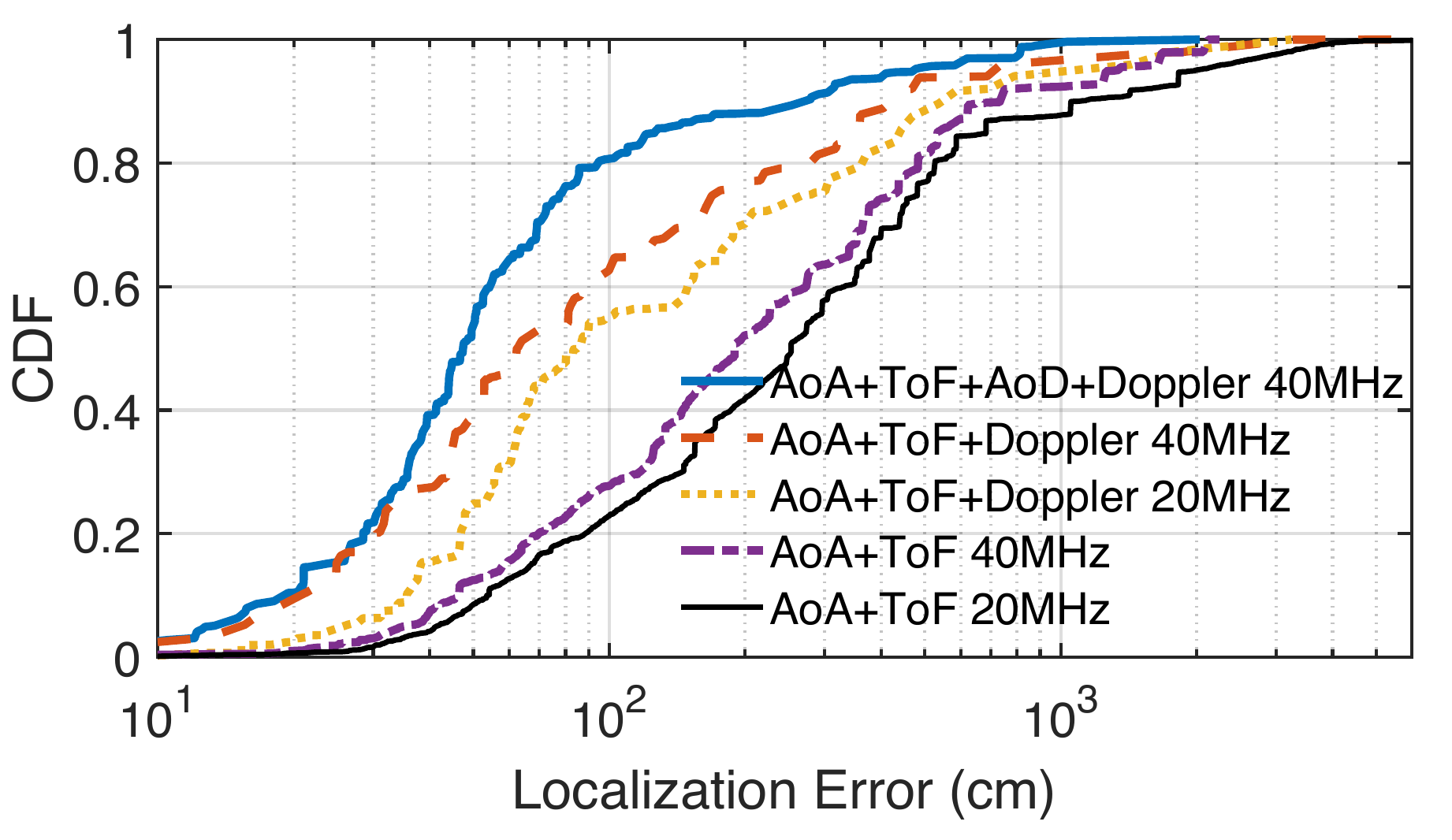}
	\caption{Localization error of \systemname when using commercial Wi-Fi devices with only 3 antennas.}
	\label{f:RouterLocErr}
		\vspace{-1em}
\end{figure}

\parahead{COTS AP performance} The results obtained using commodity Wi-Fi APs are presented in Fig.~\ref{f:RouterLocErr}. The COTS AP is equipped with only three compared with eight on the WARP platform. 
We use all three antennas of the AP but vary the size of the bandwidths and the number of dimensions of \systemname to evaluate performance. As a comparison with WARP, \systemname 3D and 4D are still able to achieve a median error of $0.67~m$ and $0.48~m$ with only three antennas.

\parahead{Impact of dimensionality} 
Our system relies on the estimated parameters to localize and track the target. Therefore, accurate parameter estimation is the premise of accurate localization and tracking. The parameter estimation is, however, severely affected by the separability or resolvability of paths. 
We evaluate how the dimensionality and resolution of each signal domain, \ie, time, angle and space, affect the resolvability, and hence the localization accuracy. 

Fig.~\ref{f:WARPlocErr} and \ref{f:RouterLocErr} depict the localization error with varying signal bandwidth~(20 and 40~MHz), antenna count~(3, 4 and 8) and dimensionality (2D, 3D and 4D). We clearly see a trend where higher numbers of dimensions lead to higher accuracy. With the same bandwidth and antenna count (\eg~8 antennas and 40~MHz), \systemname 4D achieves a much smaller median error ($0.28~m$) compared with \systemname 3D ($0.36~m$) and 2D ($1.16~m$). We see that increasing the number of antennas or bandwidth can improve the performance. By fixing the bandwidth (40~MHz) and doubling the antenna count from 4 to 8 using WARP, \systemname 3D reduces the median error from $0.57~m$ to $0.36~m$. By fixing the antenna count (3), and doubling the bandwidth from 20~MHz to 40~MHz using COST Wi-Fi AP, \systemname 3D reduces the error from $0.89~m$ to $0.67~m$. We emphasize that increasing the number of antennas~(or radio chains) and bandwidth significantly increases the hardware overhead, especially for COTS device. Increasing the dimensionality, however, does not incur any extra hardware overhead.
\begin{figure*}
	\minipage{0.32\textwidth}%
	\includegraphics[width=\linewidth]{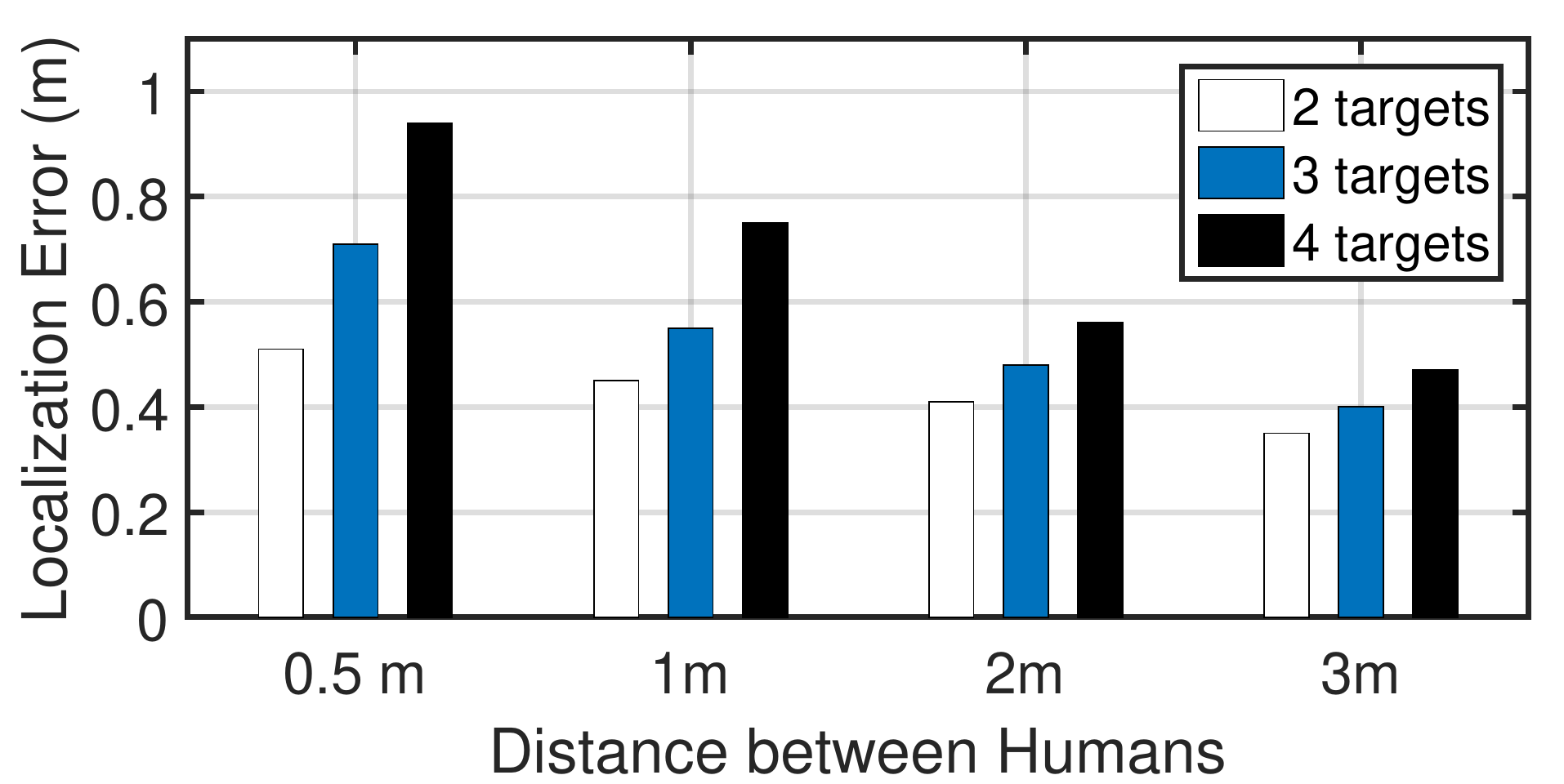}
	\caption{Localization error of  mD-Track 4D when tracking multiple
		targets separated with different distances, using WARP equipped with 8 antennas and transmitting using 40~MHz signals.}
	\label{f:multiTargetErr}
	\endminipage\hfill
	\minipage{0.32\textwidth}
	\includegraphics[width=\linewidth]{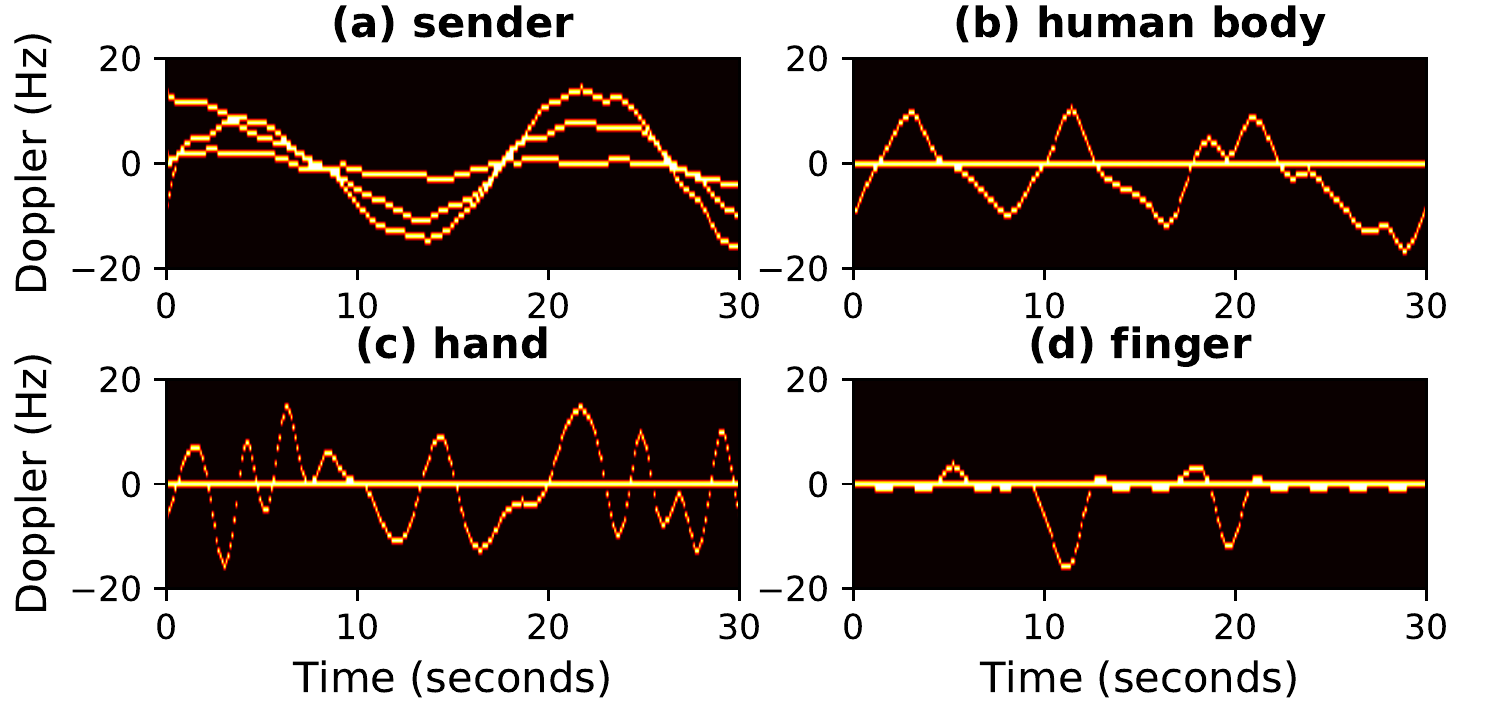}
	\caption{Doppler shifts introduced by different motions:
		\textbf{(a)} the transmitter moves; \textbf{(b)} a human walks;
		\textbf{(c)} a human stands still and waves his hand; \textbf{(d)} a
		human stands still and moves his finger.}
	\label{f:dopSpectrum}
	\endminipage\hfill
	\minipage{0.32\textwidth}
	\includegraphics[width=\linewidth]{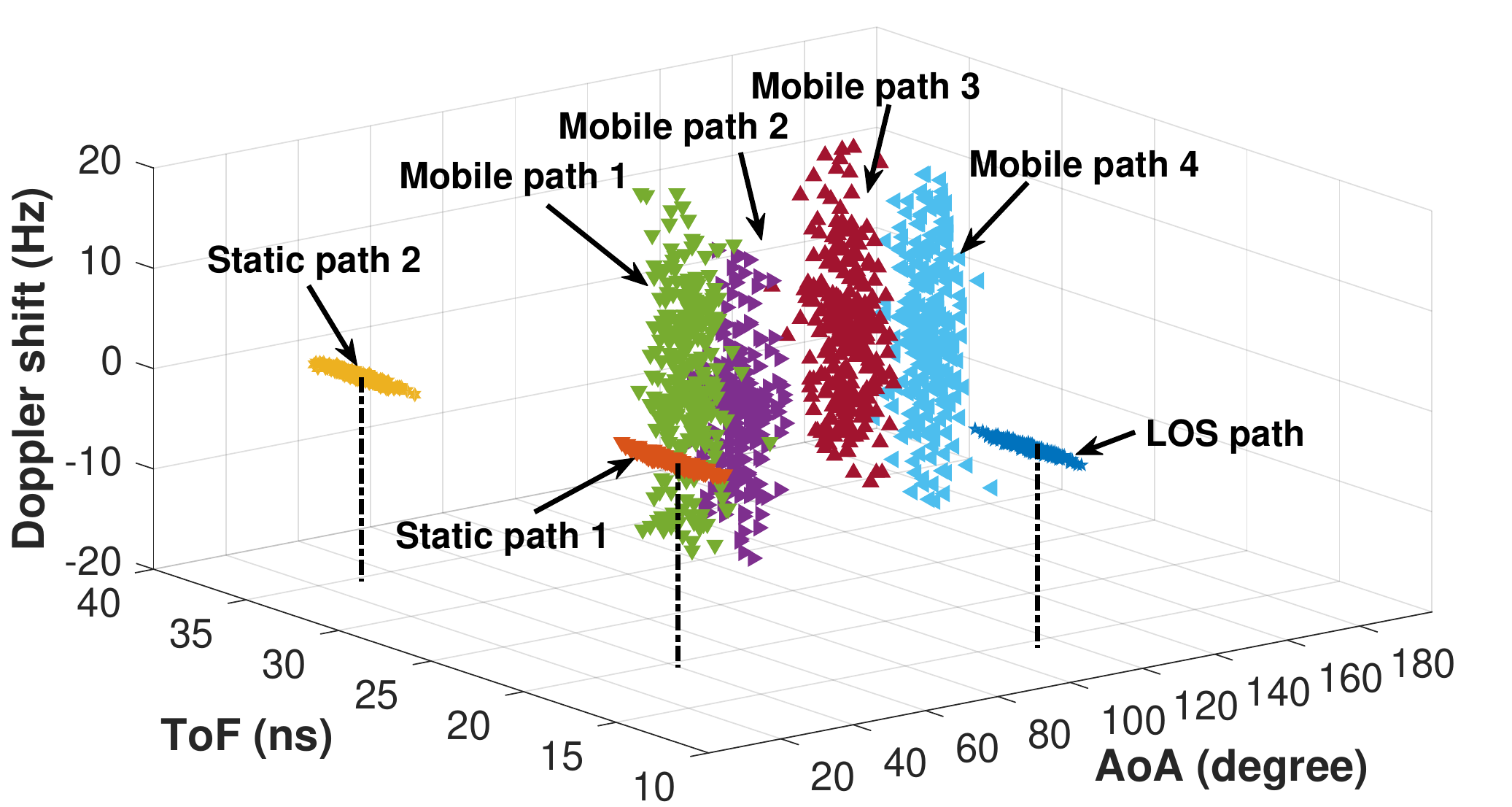}
	\caption{Estimated ToFs, AoAs and Doppler shifts. \systemname resolves three static paths in the environment and four
		mobile paths caused by four hands.}
	\label{f:estimation3d}
	\endminipage
	\vspace{-0.4cm}
\end{figure*}

\parahead{Multi-target localization} Passively localizing multiple targets in the same space is a well-known and challenging problem, especially when the targets are close to each other. The capability of resolving signals from two nearby reflectors (which we have demonstrated above), plays a key role in multi\hyp{}target localization. We ask two human to stand 0.5~m, 1~m, 2~m, and 3~m apart, and wave their hands. 
It is likely that two targets even physically close-by, have different AoA, ToF, AoD or Doppler, so that 4D \systemname is able to resolve signals from such two targets and successfully localize both of them, achieving a median error of 0.51~m.  As shown in Fig.~\ref{f:multiTargetErr}, \systemname running on the WARP platform is able to locate four targets simultaneously with a median error of 0.47~m when they are 3~m apart from each other. The error is 0.94~m even when they are 0.5~m apart.

\parahead{Detecting different types of motions}
\label{s:eval:motion}
\systemname  is not only capable of tracking the location, but also the motion of the target. In this section, we present the performance of 4D \systemname on motion tracking using WARP. We test the capability of detecting four types of motions with \systemname. In scenario one, we move the transmitter. In scenario two, the locations of the transceivers are fixed and we let a human target move towards and away from the receiver. In the third and fourth scenario, the human target stands still and only waves his hand and fingers back and forth, respectively. \systemname estimates the Doppler shift introduced by different motions and the results are depicted in Fig.~\ref{f:dopSpectrum}.
In Fig.\ref{f:dopSpectrum}~(a), we see that \systemname clearly resolves three multipath signals, and all three multipaths have non-zero Doppler shifts. This is because when the transmitter moves, all the signal paths have Doppler shifts. In Fig.~\ref{f:dopSpectrum}~(b), when the human target is moving, \systemname resolves multiple paths and only one has a non-zero Doppler shift, which is the signal reflected from the human body. Similarly, in Fig.~\ref{f:dopSpectrum}~(c), \systemname resolves multiple paths and only one has a non-zero Doppler shift caused by the hand movements. Comparing Fig.~\ref{f:dopSpectrum}~(b) and (c), we observe that the Doppler shift caused by hand movements changes more quickly than the one caused by human body movement. In Fig.~\ref{f:dopSpectrum}~(d), we can see that even the Doppler shift introduced by slight finger movements can still be detected.

\parahead{Multi-motion tracking} We demonstrate that \systemname can detect multiple the occurrence of multiple things moving. We let two people stand and wave their hands at the same time. \systemname estimates the ToF, AoA, AoD and Doppler shift of all the resolvable multipaths -- the AoA, ToF and Doppler results are shown in Fig.~\ref{f:estimation3d}. We see three static paths since their Doppler shifts are zero (one of them is the direct path).  There are four paths whose Doppler shifts are non\hyp{}zero -- their ToFs and AoAs are different from each other, which correspond to the four waving hands. The four reflected signals from the hands are clearly resolved and the parameters of each path are accurately estimated. \systemname can use Doppler shifts to detect, as well as ToF, AoA and AoD to locate multiple motions at the same time.  For each mobile path, we can use their unique time domain Doppler patterns (similar to Fig.~\ref{f:dopSpectrum}), to classify pre\hyp{}defined motions, \eg, gesture recognition~\cite{wisee-mobicom13}, activity tracking~\cite{Eeyes,WangHuman}, \textit{etc.}

\begin{figure*}[ht]%[htb]
	\centering
	\includegraphics[width=0.98\linewidth]{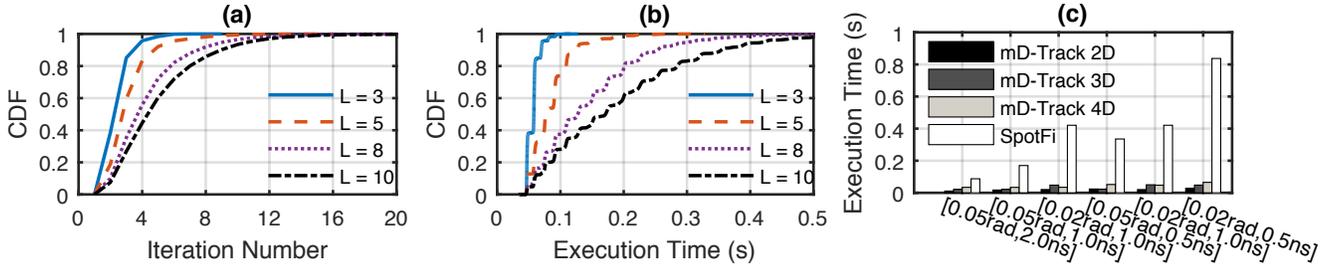}
	\caption{\textbf{(a)} Number of iterations \systemname takes to	converge with different number of multipaths. \textbf{(b)} End-to-end execution time of \systemname. \textbf{(c)} End-to-end execution time (seconds) of SpotFi and \systemname with varied granularities (step sizes).}
	\label{f:itNumRunTime}
% 	\vspace{-0.7cm}
\end{figure*}
\subsection{Computational complexity}

According to our analysis on computational complexity in Section~\ref{s:computationOverhead}, four parameters affect the computational complexity of \systemname: the number of channel paths, the number of iterations \systemname takes to converge, the dimensionality of estimation, and the number of search steps in each dimension. We run our estimation algorithm using channel measurements collected from different experimental settings and record the number of iterations, number of propagation paths, and the execution time for each trace. Fig.~\ref{f:itNumRunTime}~(a) plots the number of iterations \systemname performs before converging.  We see that \systemname also converges within four iterations 90\% of the time for traces with three dominant paths.  \systemname converges quickly within five, eight and nine iterations even when there are five, eight and ten paths, 90\% of the time.  Fig.~\ref{f:itNumRunTime}~(b) shows the overall end-to-end execution time and we see that even with eight multipaths, the median system latency is as small as 130$ms$.

Dimensionality and step size  affect the number of combinations for searching. We fix the step sizes of frequency to $0.1Hz$, vary the step size of angle and time, and evaluate the computational complexity of \systemname 2D, 3D, and 4D, in comparison with SpotFi. \systemname 2D estimates AoA and ToF jointly, as SpotFi does. \systemname conducts multiple one\hyp{}dimensional searches and thus runs much faster than SpotFi that conducts multi\hyp{}dimensional search, as depicted in Fig.~\ref{f:itNumRunTime}~(c).  In our current implementation, we use a step size of $[0.02~rad,0.5~ns,0.1~Hz]$ and as a result \systemname 2D, 3D, and 4D run 30$\times$, 20$\times$ and 14$\times$ faster than SpotFi.

\section{Related work}
\label{s:related}
%\vspace{-0.5em}
\textbf{Indoor active localization.} 
RSSI based indoor localization techniques~\cite{RADAR,Horus} provide coarse localization accuracies. More advanced techniques~\cite{Phaser,ArrayTrack} make use of an antenna array of APs~\cite{11ac1_ZhongLin,11ac_ZhongLin} to estimate the AoA of a signal. These systems can achieve accuracies of decimeters, which, however, are vulnerable to multipath propagation because of the poor resolvability due to the limited number of antennas. Moving the antenna mechanically to emulate a large antenna array is proposed~\cite{Ubicarse}, but the locations of most APs today are still fixed and the antennas cannot move freely. The latest ToF based systems~\cite{Yaxiong-mobicom15,tonetrack-mobicom15,DmLoc,Synchronicity} also provide high accuracies by combining channels to form a virtual wider bandwidth for finer resolution. However, the channel hopping those systems depend on, does affect data communication and there is only a total of 70~MHz frequency bandwidth in the 2.4GHz spectrum. Prior systems in the field tries to increase the resolution in one particular dimension. \systemname on the other hand jointly estimates multi-dimensional information to achieve a much finer resolution.
%\vspace{-0.2em}

\textbf{Passive localization and motion tracking.} A lot of attentions
has been paid to passive localization, gesture recognition and Wi-Fi
imaging in recent years. Acoustic signal based
solutions~\cite{Strata,AcuFinger,AcuFinger1} can only work within a
few decimeters. Wi-Fi RSSI/CSI signature based passive localization
systems~\cite{Eeyes,WangHuman,CSImotion1,CSImotion2,CSImotion3,MonoPHY,LiFS}
usually rely on high-density deployments which are not realistic for
large-scale deployments. Wi-Vi~\cite{WiVi} employs a signal nulling
technique to cancel signals from static objects and then estimate the
AoA of the signal reflected off the human, which can track the
direction of human's movement, but no location information can be
obtained. WiTrack~\cite{WiTrack1,WiTrack2} and other
systems~\cite{UWB1,UWB2,UWB3,UWB4,60GHz_Loc_heatherZheng,BeamSpy} use
dedicated hardware with more than 1~GHz bandwidth to achieve a high
resolution ToF for human tracking which is not possible with Wi-Fi.
Although WiTrack 2.0~\cite{WiTrack2} uses similar signal cancellation
techniques as \systemname, \systemname takes signal cancellation a
step further by integrating it with our iterative path refinement
mechanism, maximizing its efficacy. 

WiSee~\cite{wisee-mobicom13} recognizes the human gestures by estimating the Doppler shifts introduced to Wi-Fi signal by human gestures.
The key intuition of mD-Track and WiSee to address Doppler estimation is the same -- we need a long observation time interval to extract a small Doppler shift. Specifically, WiSee transmits multiple consecutive OFDM symbols (for one second) and takes a large FFT over the received OFDM symbols to estimate the small Doppler shift. Similarly, mD-Track sends multiple packets and use the received LTFs to estimate the Doppler shift. We note that \systemname not only provides motion related information, \ie, Doppler, but also estimated location related information, \ie, AoA, AoD and ToF, for every resolvable multipath.

While some recent systems~\cite{WiDeo,SpotFi,MonoLoco,WiDar,xDTrack}
consider the use of multiple parameters for object tracking, their
systems are designed to estimate fixed number of parameters, and
cannot be generalized easily to estimate parameters of more
dimensions, which \systemname enables. Furthermore, the spatial
smoothing techniques adopted by some prior work increases computational
complexity significantly \cite{SpotFi,WiDeo}

\section{Conclusion}
\label{s:concl}

\systemname is a system that incorporates information from as
many dimensions as possible to advance the accuracy of passive
wireless sensing in a multipath environment.  Our experiments
demonstrate greatly improved performance for passive multi-target
localization and motion tracking. We have also demonstrated how
\systemname can support applications including gesture recognition and
Wi-Fi imaging.

\begin{acks}
The authors would like to thank the anonymous reviewers and shepherd for their valuable comments and helpful suggestions. 
This work has been supported by Singapore MOE Tier 2 grant MOE2016-T2-2-023, Tier 1 grant 2017-T1-002-047, NTU CoE grant M4081879, and the National Science Foundation under Grant No. CNS-1617161.
	
\end{acks}
\clearpage
\bibliographystyle{abbrv}
\bibliography{paper}

\begin{thebibliography}{10}

\bibitem{80211n}
{IEEE Standard for Information technology-- Local and metropolitan area
  networks-- Specific requirements-- Part 11: Wireless LAN Medium Access
  Control (MAC)and Physical Layer (PHY) Specifications Amendment 5:
  Enhancements for Higher Throughput}.
\newblock 2009.

\bibitem{MonoPHY}
H.~Abdel-Nasser, R.~Samir, I.~Sabek, and M.~Youssef.
\newblock Monophy: Mono-stream-based device-free wlan localization via physical
  layer information.
\newblock In {\em IEEE WCNC}, 2013.

\bibitem{WiTrack2}
F.~Adib, Z.~Kabelac, and D.~Katabi.
\newblock Multi-person localization via {RF} body reflections.
\newblock In {\em NSDI}, 2015.

\bibitem{WiTrack1}
F.~Adib, Z.~Kabelac, D.~Katabi, and R.~C. Miller.
\newblock 3d tracking via body radio reflections.
\newblock In {\em NSDI}, 2014.

\bibitem{WiVi}
F.~Adib and D.~Katabi.
\newblock See through walls with {WiFi!}
\newblock In {\em ACM SIGCOMM}, 2013.

\bibitem{CSImotion2}
H.~Aly and M.~Youssef.
\newblock New insights into {WiFi-based} device-free localization.
\newblock In {\em ACM UbiComp Adjunct}, 2013.

\bibitem{RADAR}
P.~Bahl and V.~N. Padmanabhan.
\newblock {RADAR}: an in-building rf-based user location and tracking system.
\newblock In {\em IEEE INFOCOM}, 2000.

\bibitem{UWB3}
S.~Bartoletti and A.~Conti.
\newblock Passive network localization via uwb wireless sensor radars: The
  impact of {TOA} estimation.
\newblock In {\em IEEE ICUWB}, 2011.

\bibitem{EM1}
A.~P. Dempster, N.~M. Laird, and D.~B. Rubin.
\newblock Maximum likelihood from incomplete data via the {EM} algorithm.
\newblock {\em Journal of the royal statistical society. Series B}, 1977.

\bibitem{EM}
M.~Feder and E.~Weinstein.
\newblock Parameter estimation of superimposed signals using the {EM}
  algorithm.
\newblock {\em IEEE Trans.\ on Acoustics, Speech, and Signal Processing}, 1988.

\bibitem{GEM}
J.~A. Fessler and A.~O. Hero.
\newblock Space-alternating generalized expectation-maximization algorithm.
\newblock {\em IEEE Trans.\ on Signal Processing}, 1994.

\bibitem{Phaser}
J.~Gjengset, J.~Xiong, G.~McPhillips, and K.~Jamieson.
\newblock Phaser: Enabling phased array signal processing on commodity {WiFi}
  access points.
\newblock In {\em ACM MobiCom}, 2014.

\bibitem{SIC2}
D.~Halperin, T.~Anderson, and D.~Wetherall.
\newblock Taking the sting out of carrier sense: Interference cancellation for
  wireless {LANs}.
\newblock In {\em ACM MobiCom}, 2008.

\bibitem{hile-ieeegraphics08}
H.~Hile and G.~Borriello.
\newblock Positioning and orientation in indoor environments using camera
  phones.
\newblock {\em {IEEE} Computer Graphics and Applications}, 2008.

\bibitem{pharos-hotnets13}
P.~Hu, L.~Li, C.~Peng, G.~Shen, and F.~Zhao.
\newblock {Pharos: Enable} physical analytics through visible light based
  indoor localization.
\newblock In {\em HotNets}, 2013.

\bibitem{UWB2}
Y.~Jia, L.~Kong, X.~Yang, and K.~Wang.
\newblock Through-wall-radar localization for stationary human based on
  life-sign detection.
\newblock In {\em IEEE RadarCon}, 2013.

\bibitem{AngularResolution}
D.~H. Johnson and D.~E. Dudgeon.
\newblock {\em Array Signal Processing: Concepts and Techniques}.
\newblock Simon \& Schuster, 1992.

\bibitem{WiDeo}
K.~Joshi, D.~Bharadia, M.~Kotaru, and S.~Katti.
\newblock {WiDeo}: Fine-grained device-free motion tracing using {RF}
  backscatter.
\newblock In {\em NSDI}, 2015.

\bibitem{SpotFi}
M.~Kotaru, K.~Joshi, D.~Bharadia, and S.~Katti.
\newblock {SpotFi}: Decimeter level localization using {WiFi}.
\newblock In {\em ACM SIGCOMM}, 2015.

\bibitem{Ubicarse}
S.~Kumar, S.~Gil, D.~Katabi, and D.~Rus.
\newblock Accurate indoor localization with zero start-up cost.
\newblock In {\em ACM MobiCom}, 2014.

\bibitem{epsilon-nsdi14}
L.~Li, P.~Hu, G.~Shen, C.~Peng, and F.~Zhao.
\newblock {Epsilon: A} visible light based positioning system.
\newblock In {\em NSDI}, 2014.

\bibitem{VLCsense}
T.~Li, C.~An, Z.~Tian, A.~T. Campbell, and X.~Zhou.
\newblock Human sensing using visible light communication.
\newblock In {\em ACM MobiCom}, 2015.

\bibitem{UWB1}
N.~Maaref, P.~Millot, C.~Pichot, and O.~Picon.
\newblock A study of {UWB} fm-cw radar for the detection of human beings in
  motion inside a building.
\newblock {\em IEEE Trans.\ on Geoscience and Remote Sensing}, 2009.

\bibitem{GEM1}
R.~M. Neal and G.~E. Hinton.
\newblock A view of the em algorithm that justifies incremental, sparse, and
  other variants.
\newblock In {\em Learning in graphical models}, pages 355--368. Springer,
  1998.

\bibitem{cricket-mobicom00}
N.~Priyantha, A.~Chakraborty, and H.~Balakrishnan.
\newblock The {Cricket} location-support system.
\newblock In {\em MobiCom}, 2000.

\bibitem{wisee-mobicom13}
Q.~Pu, S.~Gupta, S.~Gollakota, and S.~Patel.
\newblock Whole-home gesture recognition using wireless signals.
\newblock In {\em MobiCom}, 2013.

\bibitem{WiDar}
K.~Qian, C.~Wu, Y.~Zhang, G.~Zhang, Z.~Yang, and Y.~Liu.
\newblock Widar2.0: Passive human tracking with a single wi-fi link.
\newblock In {\em ACM MobiSys}, 2018.

\bibitem{PhyCloak}
Y.~Qiao, O.~Zhang, W.~Zhou, K.~Srinivasan, and A.~Arora.
\newblock {PhyCloak}: Obfuscating sensing from communication signals.
\newblock In {\em NSDI}, 2016.

\bibitem{warp}
{Rice Univ. Wireless Open Access Research Platform ({\sc warp})}.
\newblock \url{http://warpproject.org}.

\bibitem{CUPID}
S.~Sen, J.~Lee, K.-H. Kim, and P.~Congdon.
\newblock Avoiding multipath to revive inbuilding {WiFi} localization.
\newblock In {\em MobiSys}, 2013.

\bibitem{SIC1}
S.~Sen, N.~Santhapuri, R.~R. Choudhury, and S.~Nelakuditi.
\newblock Successive interference cancellation: A back-of-the-envelope
  perspective.
\newblock In {\em ACM HotNets}, 2010.

\bibitem{ShopMiner}
L.~Shangguan, Z.~Zhou, X.~Zheng, L.~Yang, Y.~Liu, and J.~Han.
\newblock {ShopMiner}: Mining customer shopping behavior in physical clothing
  stores with {COTS RFID Devices}.
\newblock In {\em ACM SenSys}, 2015.

\bibitem{11ac1_ZhongLin}
C.~Shepard, A.~Javed, and L.~Zhong.
\newblock Control channel design for many-antenna mu-mimo.
\newblock In {\em ACM MobiCom}, 2015.

\bibitem{turbo}
B.~Sklar.
\newblock A primer on {Turbo Code} concepts.
\newblock {\em IEEE Comms.\ Magazine}, pages 94--102, Dec. 1997.

\bibitem{MonoLoco}
E.~Soltanaghaei, A.~Kalyanaraman, and K.~Whitehouse.
\newblock Multipath triangulation: Decimeter-level wifi localization and
  orientation with a single unaided receiver.
\newblock In {\em ACM MobiSys}, 2018.

\bibitem{AcuFinger}
K.~Sun, W.~Wang, A.~X. Liu, and H.~Dai.
\newblock Depth aware finger tapping on virtual displays.
\newblock In {\em ACM MobiSys}, 2018.

\bibitem{BeamSpy}
S.~Sur, X.~Zhang, P.~Ramanathan, and R.~Chandra.
\newblock Beamspy: Enabling robust 60 ghz links under blockage.
\newblock NSDI, 2016.

\bibitem{DmLoc}
D.~Vasisht, S.~Kumar, and D.~Katabi.
\newblock Decimeter-level localization with a single {WiFi} access point.
\newblock In {\em NSDI}, 2016.

\bibitem{LiFS}
J.~Wang, H.~Jiang, J.~Xiong, K.~Jamieson, X.~Chen, D.~Fang, and B.~Xie.
\newblock Lifs: Low human-effort, device-free localization with fine-grained
  subcarrier information.
\newblock In {\em ACM MobiCom}, 2016.

\bibitem{WangHuman}
W.~Wang, A.~X. Liu, M.~Shahzad, K.~Ling, and S.~Lu.
\newblock Understanding and modeling of {WiFi} signal based human activity
  recognition.
\newblock In {\em ACM MobiCom}, 2015.

\bibitem{AcuFinger1}
W.~Wang, A.~X. Liu, and K.~Sun.
\newblock Device-free gesture tracking using acoustic signals.
\newblock In {\em ACM MobiCom}, 2016.

\bibitem{Eeyes}
Y.~Wang, J.~Liu, Y.~Chen, M.~Gruteser, J.~Yang, and H.~Liu.
\newblock E-eyes: Device-free location-oriented activity identification using
  fine-grained {WiFi} signatures.
\newblock In {\em ACM MobiCom}, 2014.

\bibitem{active-badge}
R.~Want, A.~Hopper, V.~Falcao, and J.~Gibbons.
\newblock The active badge location system.
\newblock {\em {ACM Trans. on Information Systems}}, 1992.

\bibitem{bat}
A.~Ward, A.~Jones, and A.~Hopper.
\newblock A new location technique for the active office.
\newblock {\em {IEEE Personal Communications}}, 1997.

\bibitem{coordinateDescent}
S.~J. Wright.
\newblock Coordinate descent algorithms.
\newblock {\em Mathematical Programming}, 2015.

\bibitem{CSImotion1}
J.~Xiao, K.~Wu, Y.~Yi, L.~Wang, and L.~M. Ni.
\newblock {FIMD}: Fine-grained device-free motion detection.
\newblock In {\em IEEE ICPADS}, 2012.

\bibitem{Yaxiong-mobicom15}
Y.~Xie, Z.~Li, and M.~Li.
\newblock Precise power delay profiling with commodity {Wi-Fi}.
\newblock In {\em ACM MobiCom}, 2015.

\bibitem{Splicer}
Y.~Xie, Z.~Li, and M.~Li.
\newblock Precise power delay profiling with commodity wifi.
\newblock In {\em ACM MobiCom}, 2015.

\bibitem{xDTrack}
Y.~Xie, J.~Xiong, M.~Li, and K.~Jamieson.
\newblock xd-track: Leveraging multi-dimensional information for passive wi-fi
  tracking.
\newblock In {\em ACM HotWireless}, 2016.

\bibitem{SWAN}
Y.~Xie, Y.~Zhang, J.~C. Liando, and M.~Li.
\newblock Swan: Stitched wi-fi antennas.
\newblock In {\em ACM MobiCom}, 2018.

\bibitem{ArrayTrack}
J.~Xiong and K.~Jamieson.
\newblock {ArrayTrack:} a fine-grained indoor location system.
\newblock In {\em NSDI}, 2013.

\bibitem{Synchronicity}
J.~Xiong, K.~Jamieson, and K.~Sundaresan.
\newblock Synchronicity: Pushing the envelope of fine-grained localization with
  distributed {MIMO}.
\newblock In {\em ACM HotWireless}, 2014.

\bibitem{tonetrack-mobicom15}
J.~Xiong, K.~Sundaresan, and K.~Jamieson.
\newblock {ToneTrack: Leveraging} frequency-agile radios for indoor wireless
  localization.
\newblock In {\em MobiCom}, 2015.

\bibitem{Horus}
M.~Youssef and A.~Agrawala.
\newblock The horus {WLAN} location determination system.
\newblock In {\em ACM MobiSys}, 2005.

\bibitem{CSImotion3}
M.~Youssef, M.~Mah, and A.~Agrawala.
\newblock Challenges: Device-free passive localization for wireless
  environments.
\newblock In {\em ACM MobiCom}, 2007.

\bibitem{11ac_ZhongLin}
H.~Yu, O.~Bejarano, and L.~Zhong.
\newblock Combating inter-cell interference in 802.11ac-based multi-user mimo
  networks.
\newblock In {\em ACM MobiCom}, 2014.

\bibitem{Strata}
S.~Yun, Y.-C. Chen, H.~Zheng, L.~Qiu, and W.~Mao.
\newblock Strata: Fine-grained acoustic-based device-free tracking.
\newblock In {\em ACM MobiSys}, 2017.

\bibitem{UWB4}
Y.~Zhou, C.~L. Law, Y.~L. Guan, and F.~Chin.
\newblock Localization of passive target based on {UWB} backscattering range
  measurement.
\newblock In {\em IEEE ICUWB}, 2009.

\bibitem{60GHz_Loc_heatherZheng}
Y.~Zhu, Y.~Zhu, B.~Y. Zhao, and H.~Zheng.
\newblock Reusing 60ghz radios for mobile radar imaging.
\newblock In {\em ACM MobiCom}, 2015.

\end{thebibliography}
\end{document}